\documentclass[aps,showpacs,a4paper,floatfix,twocolumn,prc,amsmath,amssymb]{revtex4-1}
\usepackage{color,graphicx}
\usepackage{mathptmx}                
\usepackage{dcolumn}                 
\usepackage{bm}                      
\usepackage[utf8]{inputenc}
\def\roughly#1{\mathrel{\raise.3ex\hbox{$#1$\kern-.75em%
\lower1ex\hbox{$\sim$}}}}

\begin{document}
\title{Hyperonic stars and the symmetry energy}

\author{Constança Provid\^encia$^1$}
\author{Morgan Fortin$^2$}
\author{Helena Pais$^1$}
\author{Aziz Rabhi$^3$}
\affiliation{$^1$CFisUC, Department of Physics, University of Coimbra,
  Portugal \\
$^2$N. Copernicus Astronomical Center, Polish Academy of Sciences,
Bartycka 18, 00-716 Warszawa, Poland\\
$^3$University of Carthage, Avenue de la République BP 77 -1054
Amilcar, Tunisia}

\begin{abstract}
In the present study we analyse the effect of the density dependence
of the symmetry energy on the hyperonic content of neutron stars
within a relativistic mean field description of  stellar
matter.  For the $\Lambda$-hyperon, we consider parametrizations
calibrated to $\Lambda$-hypernuclei. For the $\Sigma$ and
$\Xi$-hyperons uncertainties that reflect the present lack of
experimental information on $\Sigma$  and $\Xi$-hypernuclei are taken into account.
We perform our study considering  nuclear equations of state that
predict two solar mass stars, and satisfy other well settled nuclear
matter properties.  The effect of the presence of hyperons on the
radius, the direct Urca processes, and the cooling of accreting neutron stars are discussed.
We show that some star properties are affected in a similar way by
 the density dependence of the symmetry energy and the hyperon content
 of the star. To disentangle these two effects it is essential to have
 a good knowledge of the equation of state at supra-saturation
 densities. 
 The density dependence of the symmetry energy affects the order of 
 appearance of the different hyperons, which may have direct
 implications on the neutron star cooling as different hyperonic neutrino processes
 processes may operate at the center of massive stars.
For models which allow for the direct Urca process to operate, hyperonic and purely nucleonic ones are shown to have a similar luminosity when hyperons are included in agreement with modern experimental data.
 It is shown that for a density dependent hadronic model constrained by experimental, theoretical and observational data,  the low-luminosity of SAX J$1808.4-3658$ can only be modelled for a hyperonic NS, suggesting that hyperons could be present in its core.
\end{abstract}

\maketitle

\section{Introduction}

The behavior of asymmetric nuclear matter is strongly influenced by the
density dependence of the symmetry energy of nuclear matter, see
\cite{Li2014} for a review. This quantity defines the properties of systems like nuclei far from the stability
line or neutron stars (NS), from the neutron skin thickness to the
NS radius \cite{Horowitz01}.
The advancement of nuclear physics and astrophysics requires,
therefore, a well-grounded knowledge of the properties of isospin-rich nuclear matter
\cite{Baran05,Steiner05,Li08}. 
 In the present study, we will concentrate
our attention on the effect of the density dependence of the symmetry energy
on some of the properties of hyperonic stellar matter that may occur
inside NSs, including  the mass and radius of hyperonic
stars \cite{Vidana10,Cavagnoli11,Providencia13} or  their cooling evolution
\cite{Prakash92,Yakovlev03}.

Although the symmetry energy is quite  well constrained  at nuclear
saturation density, see  \cite{Tsang12,Lattimer13,Oertel17}, its
density dependence at high densities is still badly known. 
The density dependence of the symmetry energy has been investigated in
many works, see for
instance
\cite{Klim07,Tsang09,Centelles09,Warda09,Carbone10,Vidana09,Ducoin11,Fattoyev14},
but usually for the saturation and sub-saturation  densities.
 Since the
description of NSs requires the knowledge of the equation of
state (EoS), from very low to very high densities, it is important to
have a correct description of the EoS in the whole range of densities.

Hyperons may have non-zero isospin, and, therefore it is expected
that the NS strangeness content and, in particular, the  non-zero
isospin 
hyperons, will be affected by the density dependence of the symmetry
energy. In the present study we will analyse the interplay between the
symmetry energy and the hyperon content in the framework of
relativistic mean-field models, following closely the work developed in
\cite{Cavagnoli11,Providencia13}, but with the care of choosing  hyperonic models
that have been calibrated to the existing experimental hypernuclei
data, as developed in \cite{Fortin17} . Besides, we will only consider
unified inner crust-core EoS since a non-unified EoS may give rise to a large
uncertainty on the star radius, as discussed in \cite{Fortin16}.

The possible existence of hyperons inside NSs has been questioned \cite{Vidana10,Demorest10}
because many of the models including hyperons  are  not able to predict
massive stars  such as the pulsars PSR J$1614-2230$
\cite{Demorest10,j1614a} and PSR J$0348+0432$ \cite{Antoniadis13} both
with a mass close to or just above two solar masses, or even the PSR
J$1903+0327$ with a mass 1.67 $\pm 0.02M_\odot$
\cite{Freire11,Vidana10}. This has been designated by the ``hyperon
puzzle'' and a review of the problem, and of the
solutions that can overcome possible contradictory scenarios has
been presented in \cite{Chatterjee15}. We will consider that the
presence of hyperons is not simply ruled out by the existence of two
solar mass stars and that this problem can be controlled by either 
using EoSs that are hard enough at high densities \cite{Fortin16} or
by going beyond the simple $SU(6)$ symmetry ansatz to fix the
isoscalar vector meson couplings \cite{Weissenborn12,Weissenborn13},
or even by considering that  nuclear matter may undergo a phase transition to quark
matter \cite{Alford06,Weissenborn11}. Having this in mind we will
explore different RMF models of nuclear matter that satisfy a set of
well-established nuclear matter properties at saturation as developed
in \cite{Fortin16}.

The paper will be organized in the following way: a review of the
formalism and presentation of the models that will be used 
in the study is given in Sec. \ref{sec:model}. In Sec
\ref{sec:icrust} and \ref{sec:hypcoup}, we discuss, respectively, the
calculation of  the inner crust EoS, and the choice of the hyperon-meson
couplings,  including the calibration of the hyperon
$\Lambda$-meson couplings for the recently proposed RMF  models %
FSU2 \cite{Chen14}, FSU2R
and FSU2H \cite{Tolos17}. In Sec. \ref{sec:esym} the effect of the
symmetry energy on the nucleonic direct Urca process, also in
the presence of hyperons, and the effect of the still-badly constrained
$\Sigma$-potential in symmetric nuclear matter on the star properties,
including cooling,
are discussed. Finally, in Sec. \ref{sec:concl} some
conclusions are drawn.

\section{The model}
\label{sec:model}
We will undertake the following discussion in the framework of a
relativistic mean field (RMF)  approach to the equation of state  of
nuclear and stellar matter. Many models have been proposed within this
framework, see the recent publication \cite{Dutra2014} for a
compilation of a large number of those models and their properties.
We will restrict ourselves to a small set, both with density
dependent couplings and non-linear meson terms, that we will justify
later. 
 Within this approach, we start from the following Lagrangian density
\begin{eqnarray}
{\mathcal L} &=& \sum_{j=1}^8  \bar \psi_j \left( i \gamma_\mu \partial^\mu - m_j +
  g_{\sigma j} \sigma
  + g_{\sigma^* j} \sigma^* \right.\nonumber \\ 
&&\left.- g_{\omega j} \gamma_\mu \omega^\mu - g_{\phi j}
\gamma_\mu \phi^\mu - g_{\rho j} \gamma_\mu \vec{\rho}^\mu
   \vec{I}_j\right) \psi_j \nonumber \\ 
&&
+  \frac{1}{2} (\partial_\mu \sigma \partial^\mu \sigma - m_\sigma^2 \sigma^2)
  - \frac{1}{3} g_2 \sigma^3 - \frac{1}{4} g_3 \sigma^4 \nonumber \\
 && 
+  \frac{1}{2} (\partial_\mu \sigma^* \partial^\mu \sigma^* - m_{\sigma^*}^2
  {\sigma^*}^2) \nonumber \\ &&
+ \frac{1}{4} c_3 (\omega_\mu \omega^\mu)^2 + {\mathcal L}_{nl}\nonumber \\ && 
- \frac{1}{4}
W_{\mu\nu} W^{\mu\nu} 
- \frac{1}{4}
P_{\mu\nu} P^{\mu\nu} 
- \frac{1}{4}
\vec{R}_{\mu\nu} \vec{R}^{\mu\nu} \nonumber \\ && 
+ \frac{1}{2} m^2_\omega \omega_\mu \omega^\mu 
+ \frac{1}{2} m^2_\phi \phi_\mu \phi^\mu 
+ \frac{1}{2} m^2_\rho \vec{\rho}_\mu \cdot \vec{\rho}^\mu ~,
\end{eqnarray}
where $\psi_j$ stands for the field of $j$  baryon, 
$\sigma, \sigma^*$ are scalar-isoscalar  meson
fields, coupling to all baryons ($\sigma$) and to strange baryons
($\sigma^*$), and $\omega^\mu$, $\phi^\mu$, $\vec\rho^\mu$ denote
the vector isoscalar (the first two)  and isovector (the last) fields,
respectively. The $\omega$ and $\vec\rho$ couple to all baryons and
the $\phi$ only to baryons with strangeness.
 $W_{\mu\nu}, P_{\mu\nu}, \vec{R}_{\mu\nu}$ are the
vector meson field tensors
$
V_{\mu\nu} = \partial_\mu V_\nu - \partial_\nu V_\mu.
$
 The couplings
$g_{\sigma N}$, $g_{\omega N}$, $g_{\rho N}$,  $g_2$, $g_3$, $c_3$,
and the  $\sigma$, $\omega$ and $\rho$ meson masses are fitted to
different kinds of data: experimental, theoretical and
observational. 
The function ${\mathcal L}_{nl}$ may be very general and defines the
density dependence of the symmetry energy. In the present study we
will limit ourselves to models with 
\begin{eqnarray}
{\mathcal L}_{nl}(\sigma,\omega_\mu \omega^\mu) &=& \left(a_1 g_\sigma^2
  \sigma^2 + b_1 g_\omega\omega_\mu \omega^\mu\right) \vec{\rho}_\mu
\cdot \vec{\rho}^\mu \nonumber\\
&=&A(\sigma,\omega^\mu\omega_\mu) \vec{\rho}_\mu\cdot \vec{\rho}^\mu,
\end{eqnarray}
where $g_{\sigma N}$ and $g_{\omega N}$ are the couplings of
the nucleons to the $\sigma$ and $\omega$ mesons. We will only consider $a_1\ne 0$ and $b_1=0$ or,
$a_1= 0$ and $b_1\ne0$. These terms have been introduced in
\cite{Horowitz01} and \cite{Horowitz01a} to explicitly model the density
dependence of the symmetry energy.

For the models with density-dependent couplings, all non-linear terms,
including the contribution ${\mathcal L}_{nl}$, are zero.
The couplings of meson $i$ to baryon $j$ are written in the form 
\begin{equation}
  g_{ij}(n_B) = g_{ij}(n_0) h_M(x)~,\quad x = n_B/n_0~,
\end{equation}
where the  density $n_0$ is the saturation density $n_0 = n_{\mathit{sat}}$ of
symmetric nuclear matter. In the present study, we consider the 
parametrizations DD2 \cite{typel10} and DDME2 \cite{ddme2}. For these
two parametrizations the functions $h_M$ assumes for the isoscalar couplings the form ~\cite{typel10},
\begin{equation}
h_M(x) = a_M \frac{1 + b_M ( x + d_M)^2}{1 + c_M (x + d_M)^2}
\end{equation}
and for the isovector couplings the form
\begin{equation}
h_M(x) = \exp[-a_M (x-1)] ~.
\end{equation}
The values of
the parameters $a_M, b_M, c_M,$ and $d_M$ can be obtained from
Ref. \cite{typel10}  for DD2 and from \cite{ddme2} for DDME2.

Both types of model with constant couplings and density-dependent couplings
will be considered in the mean field approximation, where the meson fields are replaced by their respective expectation values in uniform matter:
\begin{eqnarray}
m_\sigma^2 \bar\sigma &=& \sum_{j \in B} g_{\sigma j} n_j^s - g_2
                          \bar\sigma^2 - g_3 \bar\sigma^3 + \frac{\partial A}{\partial \bar\sigma} \bar\rho^2
\\
m_\omega^2 \bar\omega &=& \sum_{j \in B} g_{\omega j} n_j - c_3 \bar\omega^3 - \frac{\partial A}{\partial \bar\omega} \bar\rho^2 \\
m_\phi^2 \bar\phi &=& \sum_{j \in B} g_{\phi j} n_j\\
m_\rho^2 \bar\rho &=& \sum_{j \in B} g_{\rho i} t_{3 j} n_j - 2 \, A\, \bar\rho~,
\end{eqnarray}
with $\bar\rho=\langle\rho_3^0\rangle$,
$\bar\omega=\langle\omega^0\rangle$, $\bar \phi=\langle\phi^0\rangle$,
and $t_{3 j}$  the third component of isospin of baryon $j$
with the convention  $t_{3 p} = 1/2$. The scalar density of baryon
$j$ is given by
\begin{equation}
  n^s_j = \langle \bar \psi_j \psi_j \rangle = \frac{1}{\pi^2} \int_0^{k_{Fj}}
  k^2 \frac{M^*_j} dk~,
\end{equation}
and the number density by
\begin{equation}
n_j = \langle \bar \psi_j\gamma^0 \psi_j \rangle = \frac{k_{Fj}^3}{3\pi^2} ~,
\end{equation}
where 
$\epsilon_j(k) =
\sqrt{k^2 + M^{*2}_j}$, and effective chemical potential
$\mu^*_j \sqrt{k_{Fj}^2 + M^{*2}_j}$. 
The effective baryon mass $M^*_i$ is expressed in terms of the scalar mesons
\begin{equation}
M^*_i = M_i - g_{\sigma i} \bar\sigma - g_{\sigma^* i} \bar\sigma^* -
g_{\delta i} t_{3 i} \bar \delta~,
\end{equation}
where $M_i$ is the vacuum mass of the baryon $i$.
 The chemical potentials are defined by
\begin{equation}
\mu_i = \mu_i^* + g_{\omega i} \bar\omega + \frac{g_{\rho i}}{2} t_{3 i}
\bar\rho + g_{\phi i} \bar \phi + \Sigma_0^R~.
\label{mui}
\end{equation} 
where
$\Sigma_0^R$ is the rearrangement term 
\begin{eqnarray}
\Sigma_0^R &=& \sum_{j \in B} \left( \frac{\partial g_{\omega j}}\,{\partial n_j}
\bar\omega n_j + t_{3 j} \frac{\partial g_{\rho j}}\,{\partial n_j}
\bar\rho n_j +\frac{\partial g_{\phi j}}\,{\partial n_j}
\bar\phi n_j \right. \nonumber \\ && \left. -\frac{\partial g_{\sigma j}}\,{\partial n_j}
\bar\sigma n_j^s -\frac{\partial g_{\sigma^* j}}\,{\partial n_j}
\bar\sigma^* n_j^s - t_{3 j} \frac{\partial g_{\delta j}}\,{\partial n_j}
\bar\delta n_j^s \right)~,
\end{eqnarray}
and, at zero temperature, the effective chemical potential $\mu^*_i$
is given by
\begin{equation} (\mu_i^*)^2 = (M_i^*)^2 + k_{Fi}^2,\label{mueff}.
\end{equation} 
The  rearrangement  term is only present in the
density-dependent models and ensures thermodynamic consistency. 

Besides the two models with density-dependent parameters, DD2 and
DDME2, we will also consider the following  set of RMF models with
constant couplings (see Table \ref{tab:nuclear} for their properties):
FSU2 \cite{Chen14}, FSU2H and FSU2R \cite{Tolos17,Negreiros18}, NL3
\cite{nl3}, NL3 $\sigma\rho$ and NL3 $\omega\rho$
\cite{Pais16,Horowitz01}, TM1 \cite{tm1}, TM1$\omega\rho$ and
TM1$\sigma\rho$ \cite{Pais16,Bao2014}, TM1-2 and TM1-2 $\omega\rho$ \cite{Providencia13}.

\begin{table}[h!]
\center \begin{tabular}{c|cccccc}
\hline
  & $n_0$& $B$ &$K$& $E_\mathit{sym}$   &$L$ & $n_t$ \\
& $ [\mathrm{fm}^{-3}]$& [MeV]&[MeV]&[MeV]& [MeV]& $ [\mathrm{fm}^{-3}]$\\\hline 
DD2&0.149&-16.0&242.6&31.7& 55&0.067\\
DDME2&0.152&-16.1&250.9&32.3&51&0.072 \\
FSU2&0.1505&-16.28&238&37.6&113& 0.054\\
FSU2R&0.1505&-16.28&238&30.7&47 &0.083\\
FSU2H&0.1505&-16.28&238&30.5&44.5 &0.087\\
NL3 &0.148&-16.24&271&37.4&118&0.055 \\
NL3 $\sigma\rho$ &0.148&-16.24&271&31.7&55&0.080 \\
NL3$\omega\rho$&0.148&-16.24&271&31.5&55&0.081 \\
TM1 &0.145&-16.26&281&36.8&108 & 0.060\\
TM1$\omega\rho$ &0.145&-16.26&280&31.6&56&0.082 \\
TM1 $\sigma\rho$&0.145&-16.26&280&31.4&56&0.080\\
TM1-2&0.145&-16.3&281.3&36.9& 111 &0.061\\
TM1-2 $\omega\rho$&0.146&-16.3&281.7&32.1&55&0.076 \\
\hline
\end{tabular}
\caption{Nuclear matter properties of the models considered in
  this study: saturation density $n_0$, binding energy $B$,
  incompressibility $K$, symmetry energy $E_{sym}$ and its slope $L$,
  all defined at saturation density, and the crust-core transition
  density $n_t$.\label{tab:nuclear}.}
\end{table}

\section{Inner crust}
\label{sec:icrust}

In the present study we will only consider unified EoSs at the level of
the inner crust and core, since  it has been shown in \cite{Fortin16,Pais16}
that a non-unified EoS may give rise to large uncertainties in the
NS radius. The inner crust EoSs for the models we are
considering have been calculated  within the Thomas-Fermi
approximation \cite{Avancini08,Grill12,Grill14}. In the above approach, we 
assume that the inner crust is formed by non-homogeneous $npe$ matter inside a
Wigner-Seitz cell of one, two or three dimensions.  Besides, the fields are considered to vary slowly so that matter
can be treated as locally homogeneous. Since the density of the
nucleons is determined by their Fermi momenta, we can then write  the
energy as a functional of the density. The equations of motion for the
meson fields follow from variational conditions and are integrated
over the whole cell.  For a given density, the equilibrium configuration
is the one that minimizes the free energy. For the present study, we have calculated the
inner crust EoS for the models  FSU2 \cite{Chen14} and FUS2R, FSU2H
\cite{Tolos17}. In Table \ref{tab3}, we
give the density transitions between pasta configurations,
$n_{d-r}$ from droplets to rods and $n_{r-s}$ from rods to
slabs, as well as  $n_t$, the
crust-core transition density  that defines the transition  to
homogeneous matter.  $\beta$-equilibrium is imposed, and under these
conditions, the configurations corresponding to tubes and bubbles are
not present. We confirm the conclusion drawn in \cite{Oyamatsu07},
where it was discussed that models with
large values of $L$, such as FSU2, do not predict the existence of pasta
phases, due to their large neutron skin thicknesses, contrary to models with a small value of $L$, such as FSU2R
and FSU2H. As Supplementary Material 
we list the inner crust
EoS, i.e.
baryonic density, energy density and pressure, for the models FSU2,
FSU2H and FSU2R.

\begin{table}[htb]
\caption{Density transitions in the pasta phase, $n_{d-r}$ and $n_{r-s}$, for the models considered in this work. $n_t$ indicates the transition density to homogeneous matter. All densities are given in units of fm$^{-3}$. }  \label{tab3}
 \center \begin{tabular}{ccccccc}
    \hline
    \hline
 Model & \phantom{a} & $n_{d-r}$  & \phantom{a} & $n_{r-s}$  &\phantom{a} & $n_{t}$  \\
    \hline

FSU2 &\phantom{a}&   -   &\phantom{a}& -   & \phantom{a} & 0.054 \\

FSU2R &\phantom{a}& 0.037&\phantom{a}& 0.060 & \phantom{a} & 0.083 \\

FSU2H &\phantom{a}& 0.041 &\phantom{a}& 0.067 & \phantom{a} & 0.087 \\

    \hline
    \hline
  \end{tabular}
\end{table}

\section{Calibrated hyperon couplings}
\label{sec:hypcoup}

In the present study, we will only consider calibrated $\Lambda$-meson
couplings as obtained in \cite{Fortin17,Fortin18} in order to
 reproduce experimental data of $\Lambda$-hypernuclei.
The binding energies of single and double $\Lambda$-hypernuclei are
calculated solving the Dirac equations for the nucleons and $\Lambda$s,
following the approach described in ~\cite{Avancini07,Shen06}. For the
RMF models with
density-dependent couplings, we have assumed the same
density dependence for hyperon- and nucleon-meson couplings. 

Following the approach described in  \cite{Fortin17}, we have obtained
calibrated couplings for the FSU2 \cite{Chen14}, and the  FSU2R and FSU2H RMF parametrizations
recently proposed in \cite{Tolos17}. The last two parametrizations have
been fitted to both properties of nuclear matter and finite nuclei and
NS properties. The former one was fitted to ground-state properties of finite
nuclei and their monopole response. They all describe $2M_\odot$ NSs.

The values of the coupling constant fractions $R_{\sigma\Lambda}$ and  $R_{\omega\Lambda}$ to the $\sigma$ and
$\omega$ mesons are  given in
Table~\ref{tab:single}, and  $R_{\sigma^*\Lambda}$ and  $R_{\phi\Lambda}$ to the $\sigma^*$ and $\phi$ mesons in
 Table~\ref{tab:double} where $R_{\sigma\Lambda}=g_{\sigma\Lambda}/g_{\sigma N}$ and similarly for the other meson fields. For reference, we also give the $\Lambda$-potential
in symmetric nuclear matter at saturation density $n_0$ in Table~\ref{tab:single},  and  in
pure $\Lambda$-matter at $n_0$ and $n_0/5$ in  Table~\ref{tab:double} as these are quantities traditionally used to obtain hyperonic EoSs within the RMF approach.

For the coupling of the $\Lambda$ to the $\omega$ meson we
consider either the  $SU(6)$ quark model value: $R_{\omega\Lambda}(SU(6))=2/3$, the so-called models '-a', or the
maximum expected coupling, i.e.  $R_{\omega\Lambda}=1$,  forming the models '-b'.
For the coupling between the $\Lambda$ and the
$\phi$-meson  we include in the tables results obtained with the $SU(6)$ value,
$R_{\phi\Lambda}(SU(6)) =-\sqrt{2}/3$  and with $3\,R_{\phi\Lambda}(SU(6))/2=-\sqrt{2}/2$. 
We assume that the $\omega$ and $\phi$ mesons to not couple \cite{Schaffner96,Weissenborn12}.

For a given $\phi$-meson coupling, the $\sigma^*$-meson
coupling is
fitted to the bond energy of the only
double-$\Lambda$ hypernucleus for which it has been measured unambiguously, that is $_{\Lambda \Lambda}^6$He. Two sets of
parameters are given for each $\phi$ coupling corresponding to the
lower and upper values of the bond energy of  $_{\Lambda
  \Lambda}^6$He: $\Delta B_{\Lambda \Lambda}=0.50$ MeV or 0.84 MeV.

\begin{table}[h]
\center\begin{tabular}{cccc}
\hline
\hline
Model & $R_{\omega\Lambda}$ & $R_{\sigma\Lambda}$  & $U_\Lambda^N(n_0)$\\ 
\hline
FSU2-a & 2/3 & 0.619 &  -30 \\ 
FSU2-b & 1   & 0.894 &  -32 \\ 
&&&\\
FSU2R-a & 2/3 & 0.618 &  -34 \\ 
FSU2R-b & 1   & 0.893 &  -37 \\ 
&&&\\
FSU2H-a & 2/3 & 0.620 &  -35 \\ 
FSU2H-b & 1   & 0.893 &  -38 \\ 
\hline
\hline
 \end{tabular} 
\caption{Calibration to single $\Lambda$-hypernuclei: for given $R_{\omega\Lambda}$, values of $R_{\sigma\Lambda}$ calibrated to reproduce the binding energies $B_\Lambda$ of hypernuclei in the $s$ and $p$ shells. The last column contains the value of the $\Lambda$-potential in symmetric baryonic matter at saturation in MeV, for reference. }
\label{tab:single}
\end{table}

\begin{table}[h]
\center\begin{tabular}{cc|ccc|ccc}
\hline
\hline
Model &    & \multicolumn{3}{c|}{$\Delta B_{\Lambda \Lambda}=0.50$} & \multicolumn{3}{|c}{$\Delta B_{\Lambda \Lambda}=0.84$}\\
&  $R_{\phi\Lambda}$ &$R_{\sigma^\ast\Lambda}$ &  $U_\Lambda^\Lambda(n_0)$ & $U_\Lambda^\Lambda(n_0/5)$ &$R_{\sigma^{\ast}\Lambda}$& $U_\Lambda^\Lambda(n_0)$& $U_\Lambda^\Lambda(n_0/5)$ \\ 
\hline
FSU2-a & $-\sqrt{2}/3$ & 0.553 &-7.98 & -5.03 & 0.577 & -11.33 & -5.72\\ 
      & $-\sqrt{2}/2$  & 0.862 & -5.56 & -5.04 & 0.877 & -8.88  & -5.74 \\ 
FSU2-b & $-\sqrt{2}/3$ & 0.573 & 0.48  & -6.21 & 0.604 & -3.85  & -7.15 \\
      & $-\sqrt{2}/2$  & 0.874 & 5.39  & -6.18 & 0.894 & 1.15   & -7.12 \\
&&&\\

FSU2R-a & $-\sqrt{2}/3$ & 0.552 & -7.52 & -4.95 & 0.577 & -11.00 & -5.67\\ 
      & $-\sqrt{2}/2$  & 0.860 & -5.12 & -4.96 & 0.876 & -8.56  & -5.68 \\ 
FSU2R-b & $-\sqrt{2}/3$ & 0.573 & 1.31  & -6.15 & 0.604 & -3.13  & -7.11 \\
      & $-\sqrt{2}/2$  & 0.873 & 6.18  & -6.12 & 0.894 & 1.83   & -7.08 \\

&&&\\
FSU2H-a & $-\sqrt{2}/3$ & 0.544 & -8.62 & -5.52 & 0.570 & -12.16 & -6.26\\ 
      & $-\sqrt{2}/2$  & 0.848 & -6.42 & -5.53 & 0.865 & -9.93  & -6.26 \\ 
FSU2H-b & $-\sqrt{2}/3$ & 0.564 &  4.20 & -7.01 & 0.598 & -0.34  & -7.99 \\
      & $-\sqrt{2}/2$  & 0.860 &  8.75 & -6.98 & 0.883 & 4.31   & -7.96
 \\     
      \hline
\hline
\end{tabular} 
\caption{Calibration to double $\Lambda$-hypernuclei for  models -a and -b of Table \ref{tab:single}. For a given $R_{\phi\Lambda}$, $R_{\sigma^{\ast}\Lambda}$ is calibrated to reproduce either the upper or the lower values of bound energy of $_{\Lambda \Lambda}^6$He. For reference the $\Lambda$-potential in pure $\Lambda$-matter at saturation and at $n_0/5$ are also given. All energies are given in MeV.}
\label{tab:double}
\end{table}

\begin{figure}[h]
	\includegraphics[width=0.7\linewidth]{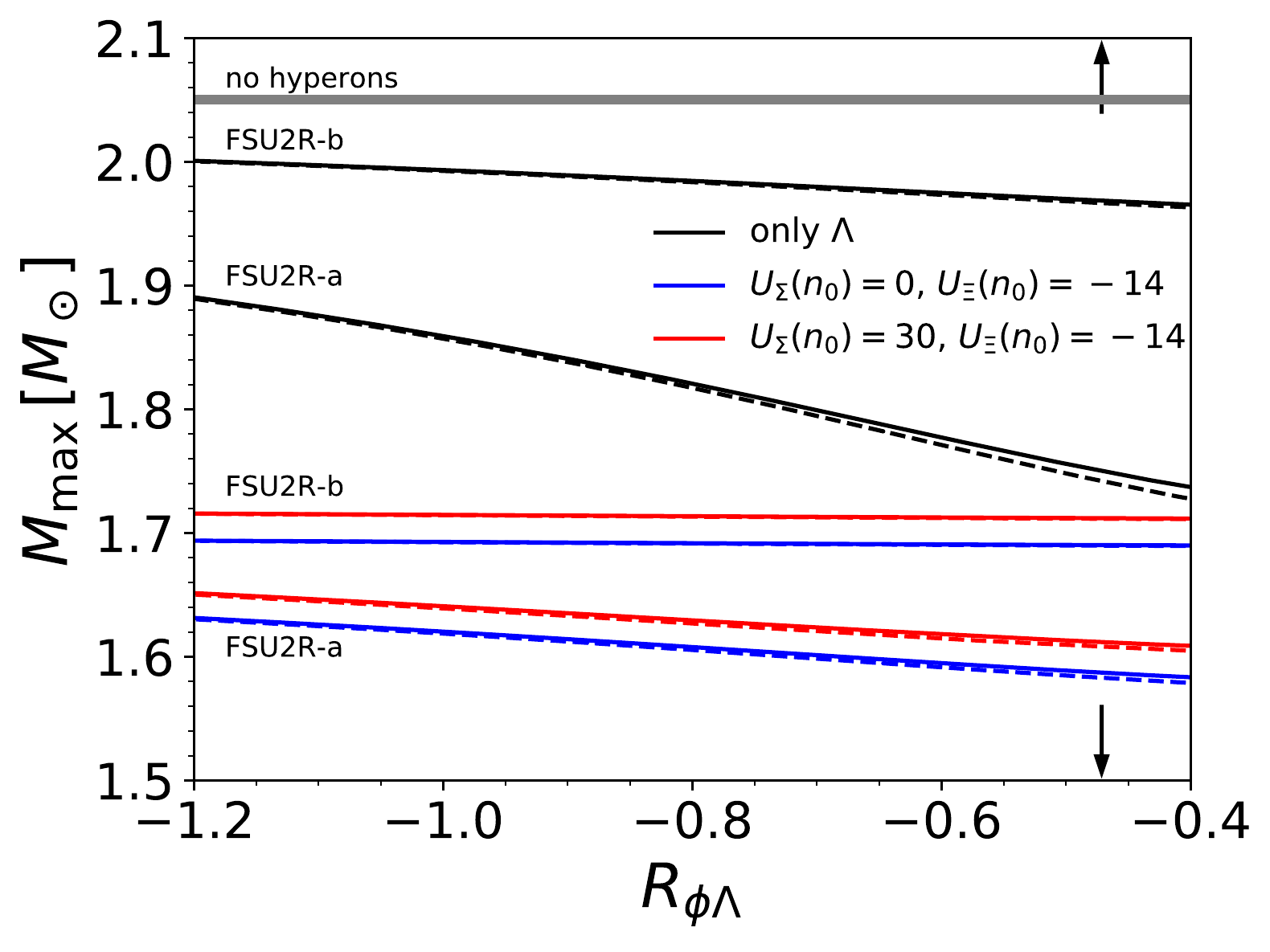}
	\includegraphics[width=0.7\linewidth]{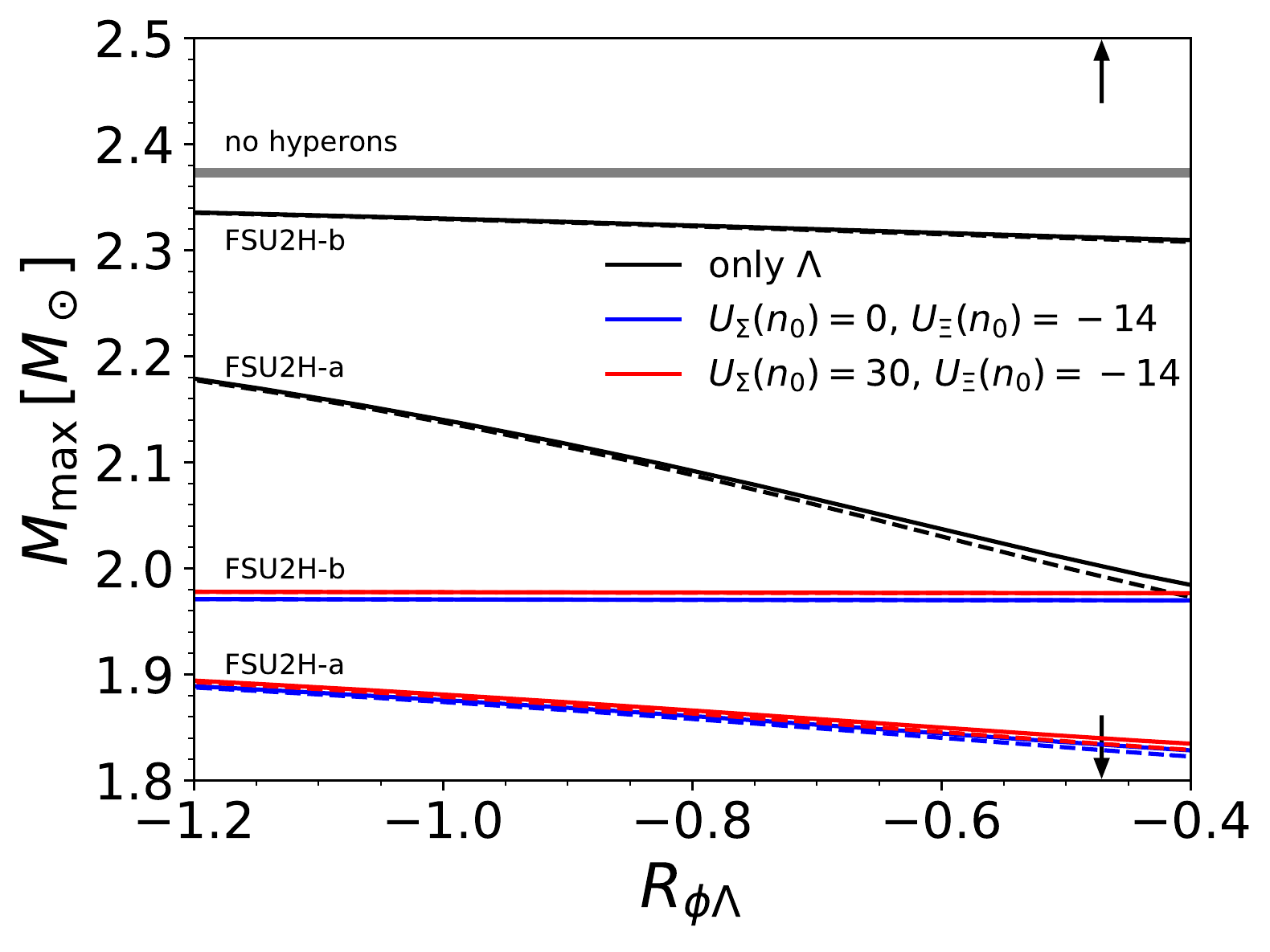}
    \caption{ NS maximum mass
      $M_{\rm max}$ as a function of $R_{\phi\Lambda}$ for FSU2R (left)
      and FSU2H (right) and
      hyperonic models -a and -b. The values
      $R_{\sigma\Lambda}$ and
      $R_{\sigma^{\ast}\Lambda}$ are adjusted to reproduce the binding
      energies of single $\Lambda$-hypernuclei and of $_{\Lambda
        \Lambda}^6$He with $\Delta B_{\Lambda \Lambda}=0.50$ MeV
      (solid lines) and 0.84 MeV (dashed lines) for chosen values of $R_{\omega\Lambda}$ and $R_{\phi\Lambda}$. The arrows indicate
      $R_{\phi\Lambda}(SU(6))$. See text for details.}
\label{mmax}
\end{figure}

To test the new parametrizations, we have integrated the
Tolman-Oppenheimer-Volkoff equations, allowing the appearance of
hyperons in the core of the star. 
  For the outer crust, we have considered  the  EoS proposed in Ref.\
  \cite{ruester06},  
and the EoS of the inner crust was obtained from a
  Thomas Fermi calculation, see \cite{Grill12,Grill14},  as discussed in the previous section, consistently with the core EoS.

With the complete EoS,  we have calculated the 
 NS maximum mass $M_{\rm max}$ as a function of
 $R_{\phi\Lambda}$ including on the $\Lambda$ hyperons in the EoS in addition to the nucleons, for the models '-a' and '-b', see black lines in Fig. \ref{mmax}. The values $R_{\sigma\Lambda}$, $R_{\phi\Lambda}$ and
 $R_{\sigma^{\ast}\Lambda}$ are adjusted to reproduce the binding
 energies of single $\Lambda$-hypernuclei and of $_{\Lambda
   \Lambda}^6$He with $\Delta B_{\Lambda \Lambda}=0.50$ MeV (solid
 lines) and 0.84 MeV (dashed lines).

 In Fig. \ref{mmax} the colored lines
 correspond to models that also include the  $\Xi$ and $\Sigma$ hyperons.
For these hyperons the values of hyperonic single-particle mean field
potentials have been used to constrain the scalar coupling constants.  The potential
for a hyperon $Y$ in symmetric nuclear matter is given by
\begin{equation}
U_Y^{N}(n_k) = M^*_Y - M_Y + \mu_Y - \mu^*_Y~,
\label{Ujk}
\end{equation}
where the chemical potential $\mu_Y$ and the effective chemical
potential $\mu^*_Y$ have   been defined in Eqs. (\ref{mui}) and  (\ref{mueff}).
For the $\Xi$ potential we take $U_\Xi^{N}(n_0)=-18$ MeV, compatible with
the analysis in \cite{khaustov00,Gal16}  of the experimental data for the reaction
 $^{12}$C$(K^-,K^+)^{12}_{\Xi^-}$Be, which are reproduced
using a potential $U_\Xi^{N}(n_0)~\sim -14$ to $-18$ MeV.
No $\Sigma$-hypernucleus has been
detected and this seems to indicate that the  $\Sigma$-potential in
nuclear matter is repulsive. Therefore, we have considered two values
of $U_\Sigma^{N}(n_0)=0$ and $+30$~MeV. 
Since, presently no information on  double $\Xi$- or $\Sigma$-
hypernuclei exists, we did not include  the coupling of these  two
hyperons to the $\sigma^*$ and the $\phi$-meson, responsible for the
description of the $YY$ interaction in RMF models. For the
$\omega$-meson couplings we consider the $SU(6)$ values:
 \begin{equation}
g_{\omega\Xi}=\frac{1}{3} g_{\omega N} = \frac{1}{2} g_{\omega\Sigma} .
\end{equation}

\begin{figure}[h]
\includegraphics[width=0.9\linewidth]{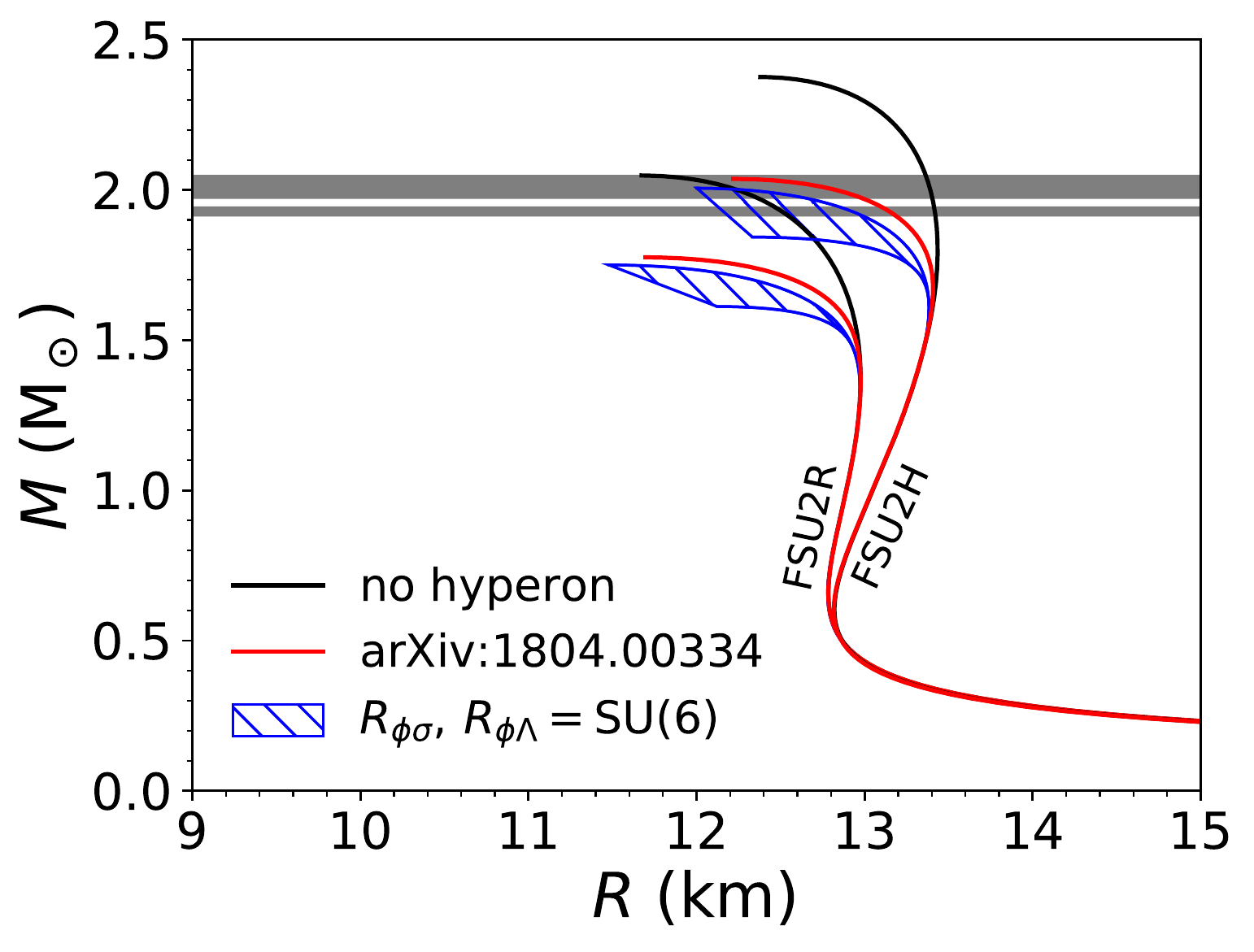}
\caption{$M-R$ relations for the FSU2R and FSU2H models. The grey
  strips correspond to the mass of the two heaviest known NSs, PSR
J$1614-2230$ and PSR J$0348+0432$. The black
  lines are obtained for purely nucleonic models, the red ones for the
  models presented in \cite{Negreiros18}. The blue
  stripped areas correspond to the mass range covered when employing
  the minimal and maximal hyperonic models for SU(6) values of the
  coupling constants $R_{\omega\Lambda}$ and $R_{\phi\Lambda}$ of the $\Lambda$
  hyperons, for $U_\Sigma(n_0)=0$ MeV and $U_\Xi(n_0)=-14$ MeV - see
  Fig. \ref{mmax}.}
\label{mr}
\end{figure}

In Fig. \ref{mmax} the predictions obtained with the EoSs that include only the $\Lambda$ hyperons in addition to the nucleons defining the minimal hyperonic model  (black lines),  may be considered as an
upper limit on the maximum mass of an hyperonic NS, when
compared with
models including the full baryonic octet. On the other hand, including
in the calculations
the complete baryonic octet  and not including
the mesons that account for the YY interaction (colored lines), the maximal hyperonic model,  gives an
estimation of the lower limits for the maximum mass of hyperonic NSs. 
The blue
  stripped areas  in Fig. \ref{mr} correspond, precisely, to the mass range covered when employing
  the minimal and maximal hyperonic models for $SU(6)$ values of the
  coupling constants $R_{\omega\Lambda}$ and $R_{\phi\Lambda}$ of the $\Lambda$
  hyperons, for $U_\Sigma(n_0)=0$ MeV and $U_\Xi(n_0)=-14$~MeV.

Under the above conditions the FSU2R model with hyperons does not
 describe two solar mass stars (not even 1.9 $M_\odot$ as
indicated by the most recent measurements of PSR J$1614-2230$ \cite{j1614a}). This
conclusion had already been drawn in \cite{Tolos17}. In Fig. \ref{mr} the red curves have
been obtained with the hyperon parametrization defined in
\cite{Tolos17}. It lies above the upper limit defined by the
minimal hyperonic model because the $\sigma^*$ was not included,  and the $\Lambda$-$\sigma$
coupling was also smaller giving rise to a potential equal to -28 MeV
instead of  $\sim -35$ MeV  obtained with the calibrated parametrization.

\section{Symmetry energy and hyperonic neutron stars}
\label{sec:esym}

In the present section, we discuss the effect of the density
dependence of the symmetry energy on the onset of the different
hyperon species, and on the onset of the direct Urca process in the presence of
hyperons.
The study will be undertaken considering a family of models generated
from the TM1 model \cite{tm1}. The inclusion of the  nonlinear term  ${\mathcal L}_{nl}$ that
couples the $\omega$ and the $\sigma$ mesons to the  $\rho$-meson
will allow the  generation of a family of
models with the same underlying isoscalar properties and different
isovector properties \cite{Bao2014,Pais16}. 
 This family is built in
such a way that all the models predict the same symmetry energy, 
equal to the one predicted by TM1, at $n_B=0.1$ fm$^{-3}$. It was
shown in \cite{Pais16} that the ground-state properties of nuclei used
to calibrate TM1 are still quite well reproduced when the new terms
are introduced in the model. Contrary to the previous section, in the
present and following sections we will consider that the $\Sigma$ and
$\Xi$ hyperons couple to the $\phi$-meson with the couplings defined
by the $SU(6)$ symmetry,  unless when Fig. \ref{uxi}
is discussed.

\subsection{The direct  Urca process: nucleonic neutron stars}

\begin{figure}[t]
\begin{center}
\begin{tabular}{c}
\includegraphics[angle=0, width=1.\linewidth]{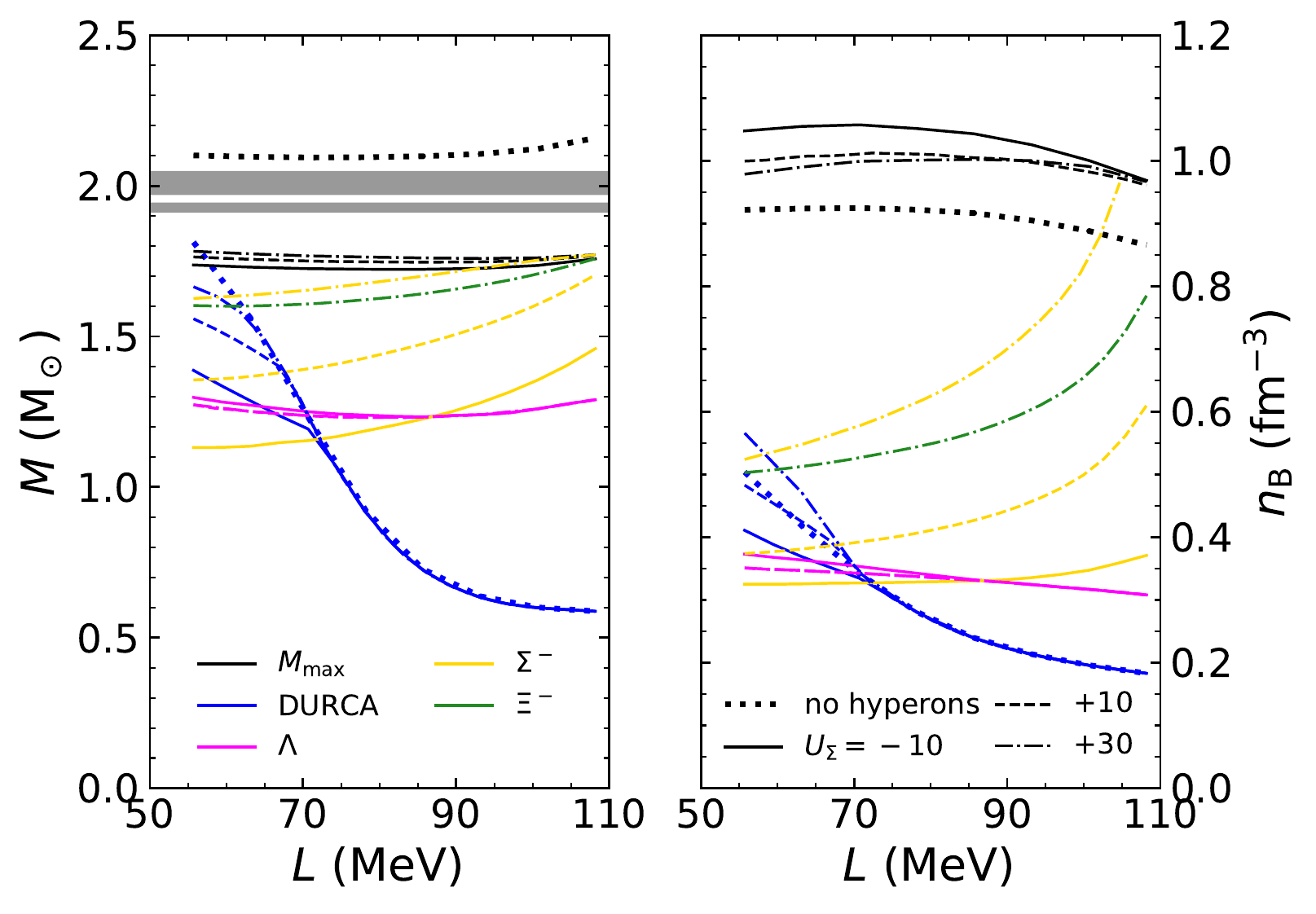}
\end{tabular}
\end{center}
\caption{Right panel: Onset density of the different hyperons
  ($\Lambda$, $\Sigma^-$ and $\Xi^-$) (red, yellow and green lines),
  and of the DU process (blue lines), and
  NS central density at the maximum mass  (black lines) for the TM1$\omega\rho$ family and three different values of the $\Sigma$ potential at
  saturation ($-10$, 10, and 30 MeV) as a function of the slope of the symmetry
  energy $L$. Left panel: the NS
  masses corresponding to the different densities plotted in the right panel.  The DU onset density  in nucleonic matter and
  corresponding star mass are also shown with blue dots. 
All other curves were obtained with models
  including hyperons. }
\label{fig:tm1}
\end{figure}

The most  efficient cooling mechanism of a NS  by neutrino emission is the
nucleonic electron
direct Urca (DU) process \cite{DU91} described by the equations
\begin{equation}
n \rightarrow p + e^- + \bar{\nu}_e  \qquad \textrm{and} \qquad p+e^- \rightarrow n+\nu_e.
\label{eqn:DUrcanpe}
\end{equation}
This process operates only if momentum conservation is allowed,
and this can be translated into the inequalities:
\begin{equation}
p_{{\rm F}n}\leq p_{{\rm F}p} + p_{{\rm F}e},
 \label{eq:DUrcaN}
\end{equation}
where $p_{{\rm F}i}$ is the Fermi momentum of  species $i$. As a
consequence, in order for the DU process to occur the proton fraction
must be equal or above  a minimum proton fraction $Y_p^{\rm min}$ \cite{Klahn06}: 
\begin{equation}
Y_{p}^{\rm min}=\frac{1}{1+\left(1+x_e^{1/3}\right)^{3}},
\end{equation}
where $x_e=n_e/\left(n_e+n_\mu\right)$, and $n_e$ and  $n_\mu$ are  the
electron and muon  densities.  In the following, we will designate by
$n_{\rm DU}$ and mass $M_{\rm DU}$, respectively, the baryonic density
at which the DU process sets in and the mass of the star where it starts
operating, i.e. which has a central density equal to $n_{\rm DU}$.

For some models the nucleonic DU process does not operate
inside NSs because the  onset DU density is above the central
density of the most massive star. In our study this is the case for the two models
with density-dependent  coupling parameters DD2 and DDME2.

In order to discuss the influence of the density dependence of the
symmetry energy on the DU process, we include in Fig. \ref{fig:tm1} left
panel the DU
onset density as a function of the slope $L$ of the symmetry
energy at  saturation density (blue curves) and the corresponding star
masses on the right panels. The blue dotted line is obtained for the
nucleonic EoSs from the family of TM1 models defined in
section \ref{sec:model} and the other blue curves have been obtained for
hyperonic EoSs and will be discussed below.  It is clear that the DU
process is strongly influenced by the density dependence of the
symmetry energy, because this quantity  defines the proton
fraction in matter. A similar relation was obtained in
\cite{Cavagnoli11,Providencia14}.
 A large symmetry energy disfavors a large
proton-neutron asymmetry and, therefore, favors the DU process and it sets in at low densities. On the
contrary, a small symmetry energy allows for large proton-neutron
asymmetries hence pushing the DU threshold to higher densities. In \cite{Horowitz02}, the authors have discussed how it
is possible to establish a relation between the $^{208}$Pb  neutron
skin  and the possibility of occurring the DU process. Since  the nuclear  neutron skin is
strongly correlated with the slope $L$, the above observation is
equivalent to the one displayed in Fig. \ref{fig:tm1}.

\subsection{The direct Urca process: hyperonic neutron stars}
\label{sec:hyp}

In the presence of hyperons, other channels are opened for neutrino
emission \cite{Prakash92}: 
\begin{eqnarray}
\Sigma^-&\to&\Sigma^0\ell^- \bar \nu_\ell,\,\quad R=0.61 \label{y2}\\
\Xi^-&\to& \Xi^0 \ell^- \bar \nu_\ell,\,\quad R=0.22 \label{y5}\\
\Sigma^-&\to& \Lambda \ell^- \bar \nu_\ell,\,\quad R=0.21 \label{y3}\\
\Xi^0&\to& \Sigma^+ \ell^- \bar \nu_\ell,\,\quad R=0.06 \label{y7}\\
\Lambda\, &\to& p \ell^- \bar \nu_\ell,\,\quad R=0.04 \label{y1}\\ 
\Xi^-&\to& \Sigma^0 \ell^- \bar \nu_\ell,\,\quad R=0.03 \label{y8}\\
\Xi^-&\to& \Lambda \ell^- \bar \nu_\ell,\,\quad R=0.02 \label{y6}\\
\Sigma^-&\to&n\ell^- \bar \nu_\ell,\,\quad R=0.01 \label{y4}.
\end{eqnarray}
For each process the $R$ factor indicates the efficiency of each process with respect to the nucleonic DU process for which $R=1$ (see \cite{Prakash92}). These different hyperonic DU channels are opened as soon as the species involved set in. The most
efficient processes being the ones described by Eqs.
(\ref{y2}), (\ref{y5}) and (\ref{y3}) and, in particular, the process
(\ref{y2}) is almost three times  more efficient that the other
two. This indicates that it is important to establish whether the
$\Sigma$-hyperon occurs inside a NS. Since this hyperon has isospin
equal to one, it is expected that its occurrence will be strongly
influenced by the density dependence of the symmetry energy.

\begin{figure}[ht]
\begin{center}
\begin{tabular}{c}
\includegraphics[angle=0, width=0.9\linewidth]{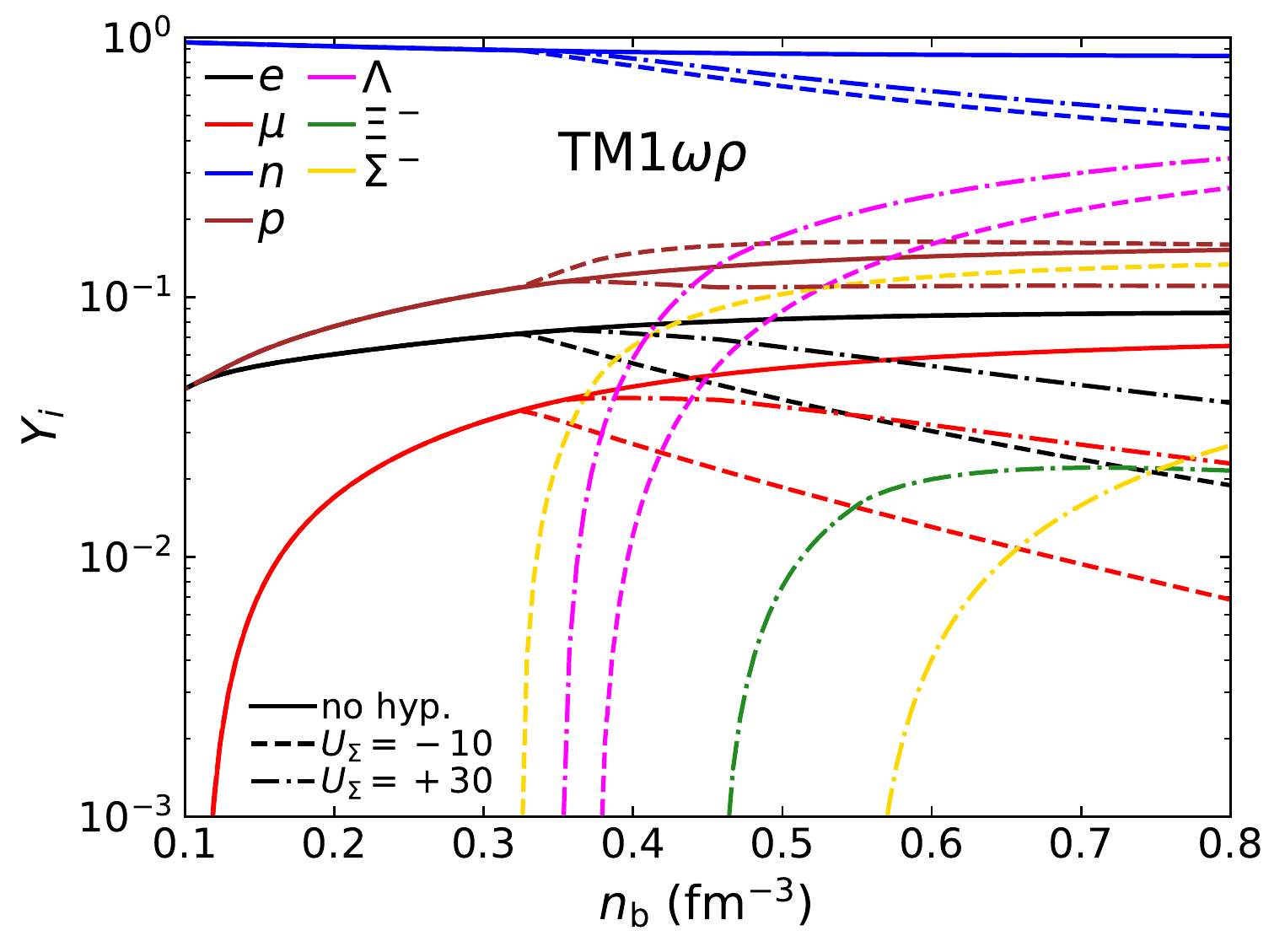}\\
\end{tabular}
\end{center}
\caption{Particle fractions for $U_\Sigma=+30$ MeV (black lines
  with marks), $U_\Sigma=-10$ MeV (color lines
  with marks), $npe\mu$ matter (thin black lines), obtained with model
  TM1$\omega\rho$.}
\label{fig:frac}
\end{figure}

The occurrence of hyperons affects the neutron, proton and electron
fractions. Therefore, 
Eq. (\ref{eqn:DUrcanpe}) for the nucleonic DU threshold looses validity, and after hyperons set
in, the minimum proton
fraction for nucleonic electron DU is given by
\begin{equation}
\left(\frac{n_{p}}{n_p+n_n}\right)=\frac{1}{1+\left(1+{x^Y_e}^{1/3}\right)^{3}}, \quad x_e^Y=\frac{n_e}{n_e+n_\mu-n_Y^{ch}},
\label{DUrcaYpn}
\end{equation}
where $n_Y^{ch}=-n_{\Sigma^-}+n_{\Sigma^+}-n_{\Xi^-}$. The nucleonic
electron DU process is not affected by the presence of hyperons in
models with a large slope $L$ because its threshold is at densities lower than the hyperon
onset density. However, if $L\le 75$ MeV, the presence of hyperons will
affect the nucleonic electron DU process and the effect depends on the
value of the $\Sigma$ potential: if very repulsive ($U_\Sigma$ of the order of couple of tens of MeV), the DU process turns on
 at densities larger that the one obtained for nucleonic
matter. The contrary holds for less repulsive $\Sigma$  potentials.

In Fig. \ref{fig:frac}, the fractions of the particles present inside a NS
star below $n=0.8$ fm$^{-3}$ for the TM1$\omega\rho$ parametrization are shown for hyperon free matter (thin
black lines) and for hyperonic matter taking $U_\Sigma (n_0)=-10$ and
+30 MeV.  For the attractive potential ($U_\Sigma$ at saturation negative) the $\Sigma^-$ is the first
hyperon to set in and as soon as it appears the proton fraction
increases and the neutron fraction decreases, reducing the difference
between the proton and neutron Fermi momenta and favoring the DU
process relative to nucleonic matter. For the very repulsive
potential at saturation: $U_\Sigma=30$ MeV, a value that is generally employed in the recent literature, the $\Lambda$ is the first hyperon to set in and above its densities of appearance the
fractions of neutrons, protons, electrons and muons all suffer a
reduction, the overall effect being that DU is disfavoured with
respect to nucleonic matter.

In Fig. \ref{fig:tm1} left panel, which was partially discussed before, we also
plot, besides the onset density of
the nucleonic electron DU process,  the onset densities of
the $\Lambda$, $\Sigma^-$ and $\Xi^-$ hyperons, and the central density $n_c$ of the NS with the maximum mass for three different values of $U_\Sigma$ at saturation: $-10$, 10 and 30 MeV. Hyperons that are not included in the figure do not appear at densities 
below $n_c$ and hence are not present at all in NSs. The grey bands show the mass constraints set by the pulsars PSR
J$1614-2230$ and PSR J$0348+0432$. Even though the  TM1$\omega\rho$ family with hyperons and the vector
meson couplings to the hyperons defined by the $SU(6)$ symmetry do not satisfy
the two solar mass constraint, the main conclusions drawn with respect to the $L$
dependence of the several properties we  discuss, is still valid
for more massive stars. 

For $L\ge 75$ MeV the DU process sets in at a density below the
hyperon onset density and, in fact, the DU process is possible at
densities of the order of 2$n_0$ or below, corresponding to stars with
a mass equal to 1$M_\odot$ or below.  Observations do not support a
fast cooling for these low masses (see eg. discussion in \cite{Fortin18a}). The DU mass threshold rises monotonously as $L$ decreases below $ 75$ MeV, and for
$L=50$ MeV attains  $1.4-1.7
M_\odot$ depending on the value of $U_\Sigma$, a large repulsive
value favoring a higher threshold. Similar conclusions have been
drawn in \cite{Cavagnoli11}, although using different hyperonic models.

We finally comment on the effect of $L$ on the hyperonic species
inside the star. The $\Lambda$ hyperon onset is practically not affected by
the value of $U_\Sigma$, and, although its onset density increases
slightly when $L$ decreases, the mass of star at the $\Lambda$-onset
is essentially independent of $L$ and equal to 1.3$M_\odot$.  However,
the other two hyperons $\Sigma^-$ and $\Xi^-$, having a non-zero isospin
are strongly affected by the density dependence of the symmetry
energy, the onset density decreasing as $L$ decreases. The more
repulsive the $U_\Sigma$  the larger the onset density of the $\Sigma$ and the
mass of the star where the hyperon sets in.

The strongest effect of the $U_\Sigma(n_0)$ is observed for $L=56$ MeV. 
In nucleonic matter the DU sets in at $n_{DU}=0.504$ fm$^{-3}$
corresponding to a star with a mass $M_{DU}=1.81\, M_\odot$. The
density $n_{DU}$ and mass $M_{DU}$ change to  $n_{DU}=0.411$ fm$^{-3}$
and $M_{DU}=1.39\, M_\odot$ if $U_\Sigma(n_0)=-10$ MeV, and
to   $n_{DU}=0.566$ fm$^{-3}$
and $M_{DU}=1.67\, M_\odot$ if $U_\Sigma(n_0)=+30$ MeV,
The $\Xi^-$ does not occur unless the $\Sigma$
potential is quite repulsive.

One fact that should be pointed out  is that
the overall effect of the value of  $L$ on the star maximum
mass is negligible, a conclusion that had already been drawn in \cite{Cavagnoli11,Providencia13}.

\begin{figure}[htb]
\begin{center}
\begin{tabular}{c}
\includegraphics[angle=0, width=1.\linewidth]{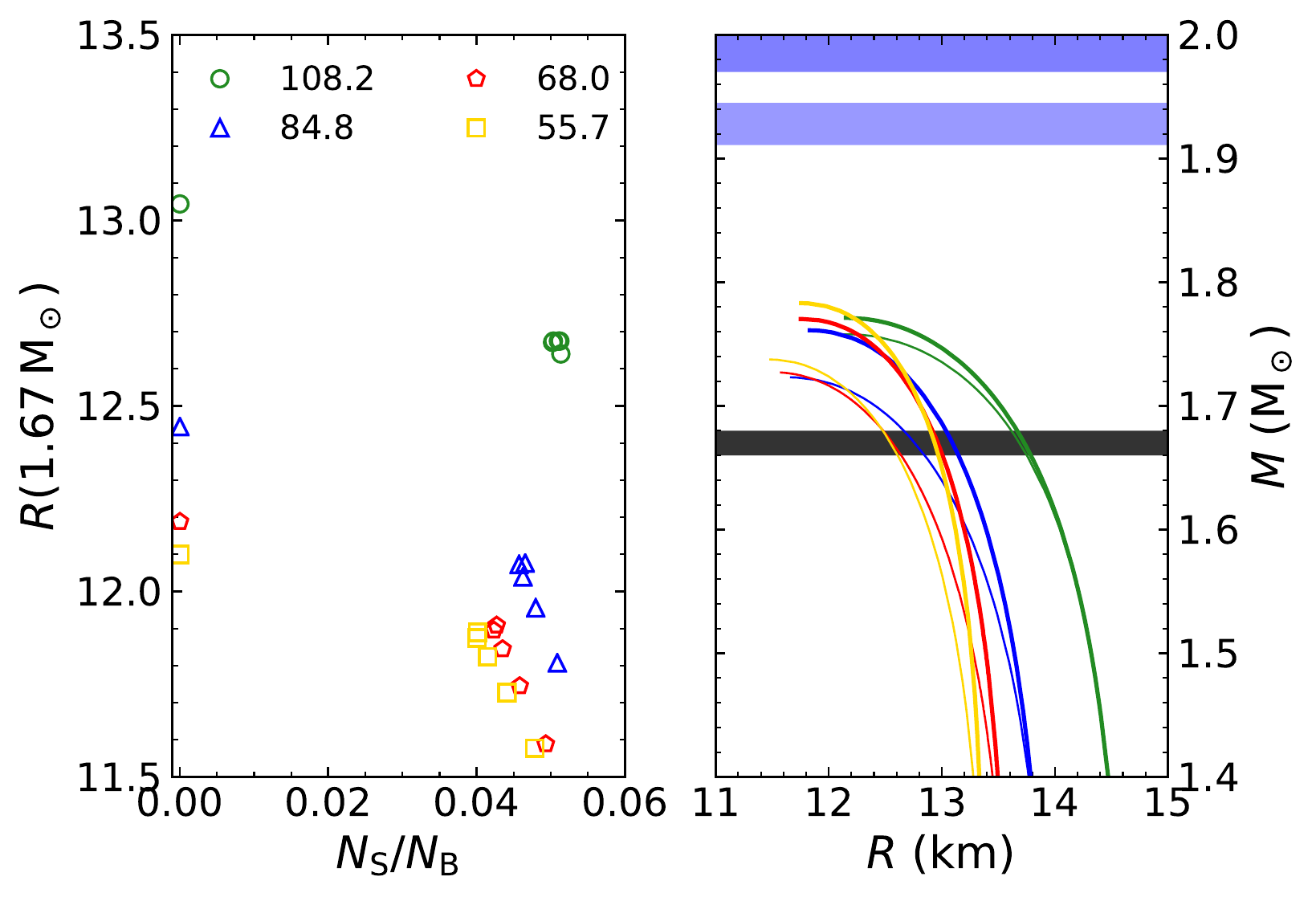}\\
\end{tabular}
\end{center}
\caption{Left panel: Radius of a 1.67$M_\odot$ NS as a function of the
  total hyperon fraction $N_{\rm S}$ that was normalized by the total
    baryon number $N_{\rm B}$ at the maximum mass for four  TM1$\omega\rho$ models with 
    four different values of the slope $L$: 55.7, 68.0, 84.8 and 108.2   MeV. Right panel: M-R curves obtained with the models with the  different  values of $L$  for  $U_\Sigma=+30$ MeV (thick lines) and $-10$ MeV (thin lines). The line color indicates the value of the $L$ as in the left panel. The blue
 upper bands show the constraints set by the pulsars PSR J$1614-2230$ and PSR J$0348+0432$.}
\label{rnstm1}
\end{figure}

In Fig. \ref{rnstm1} left panel, we show how the radius of NSs
with a mass equal to 1.67$M_\odot$, the mass of the pulsar  PSR
J1903+0327,  changes with  the total hyperon fraction, one
third of the strangeness fraction when only hyperons with strangeness
charge -1 are involved,  at the maximum
mass 
$$ N_S=\frac{1}{3}\int_0^R  dr\frac{n_s r^2}{\sqrt{1-m(r)/r}},
$$
where $m(r)$ is the mass inside the radius $r$ and $n_s$ the
strangeness density, obtained when
$U_\Sigma$ varies between $-10$ and +30 MeV and the other hyperon coupling parameters are
kept unchanged for parametrizations of the TM1$\omega\rho$ family with
different values of $L$. 

 The right panel of the same figure represents the M-R
curves of the same models.
 For $L=108$ MeV the $\Lambda$-hyperon is  the
responsible for almost all the  strangeness content and, therefore, it
is not sensitive to the $\Sigma$ potential. On the other hand, 
models with smaller
values of $L$ are sensitive to the $\Sigma$ potential and a change of
$U_\Sigma(n_0)$  between $-10$ and +30 MeV is translated into a reduction of
$\sim20\%$ of the total strangeness content and an increase of $300-400$ m of
the star radius.  
The overall effect  on the radius due to the
inclusion of hyperons in the
family of models considered in this section is a
reduction of at most 400 to 600m.
Let us recall that several
authors, including
\cite{Horowitz01a,Horowitz01a,Lattimer01,Carriere03,Cavagnoli11,Lattimer14,Pais16},
have shown that  the NS radius is correlated with the nucleus neutron
skin, a quantity directly related with the slope of the symmetry
energy: the larger the slope of the symmetry energy the larger the
radius. This behavior is clearly seen in the left panel of
Fig. \ref{rnstm1} : for the non-hyperonic models, located on the vertical axis where $N_S/N_B=0$ of the left panel, the radius of a $1.67M_\odot$ increases with the symmetry energy slope $L$, and a
difference in radius of almost 1 km is obtained between models with $L=56$ MeV and $L=108$ MeV.

\subsection{Effect of the $\Sigma$ potential}

It was shown in the previous section that besides the symmetry energy
the value of $\Sigma$ potential in symmetric matter at saturation, chosen to fix
the value of the $\sigma$-meson coupling, could also have a strong
effect on the properties of the star, in particular, if the model has a small value of $L$.
In the following, we analyse this effect and,
taking into account that  the $\Sigma$-meson
interaction is still  not constrained, we  allow it to vary between $-10$ MeV
and 30 MeV. Experimentally no $\Sigma$-hypernucleus was detected and
this seems to indicate that the  $\Sigma$ interaction in nuclear
matter is repulsive or, at most, slightly attractive.

\subsubsection{Direct Urca process}

\begin{figure}[ht]
\begin{center}
\begin{tabular}{c}
\includegraphics[angle=0, width=0.95\linewidth]{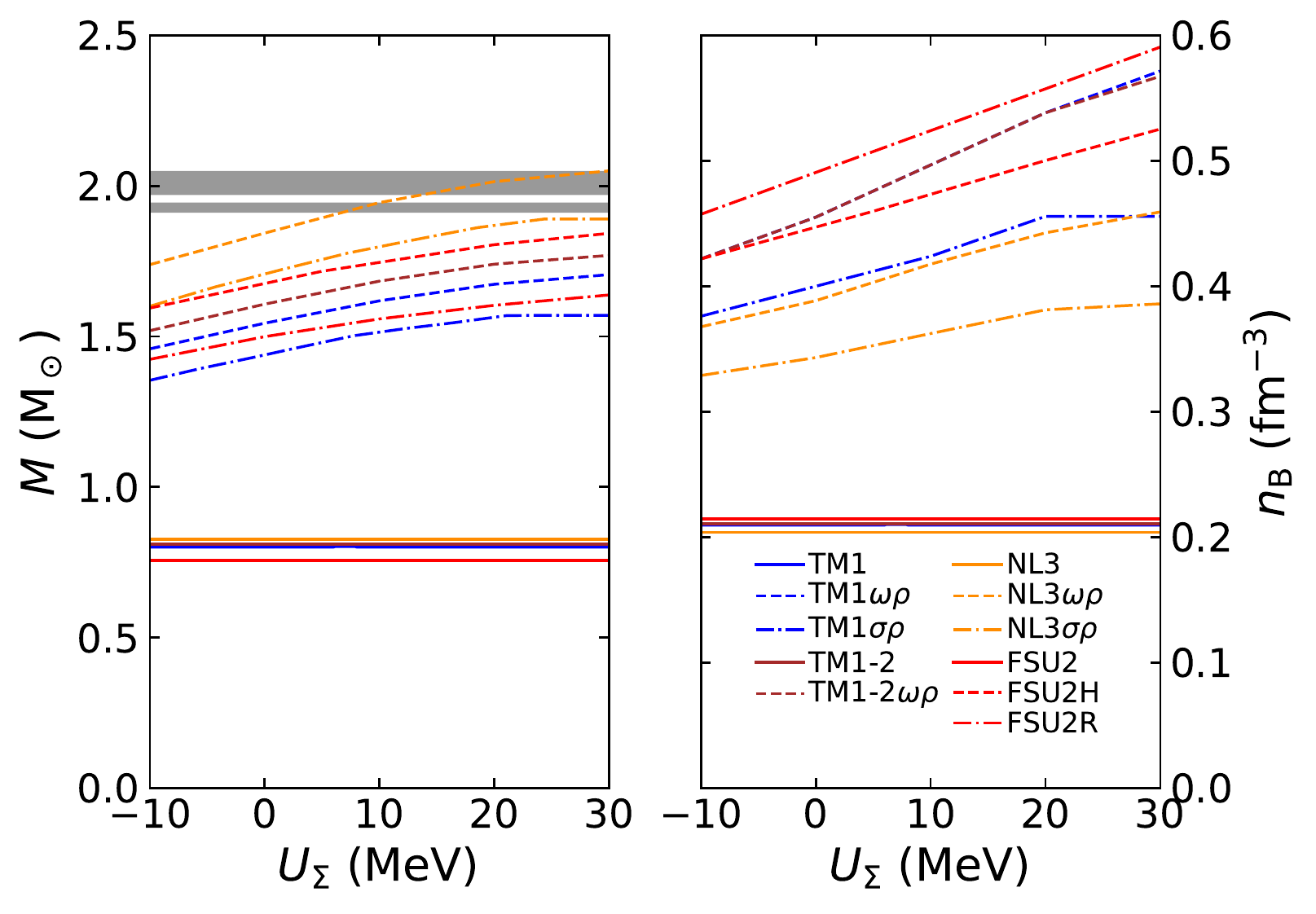}
\end{tabular}
\end{center}
\caption{Left panel: Onset density  $n_{\rm DU}$ of the DU process for models in which the DU process turns on in NSs, as a function of the
  $\Sigma$ potential at saturation density. Right panel: corresponding
  NS masses. All models considered
  include hyperons.}
\label{sigma}
\end{figure}

\begin{figure}[t]
\begin{center}
\begin{tabular}{c}
\includegraphics[angle=0, width=0.45\textwidth]{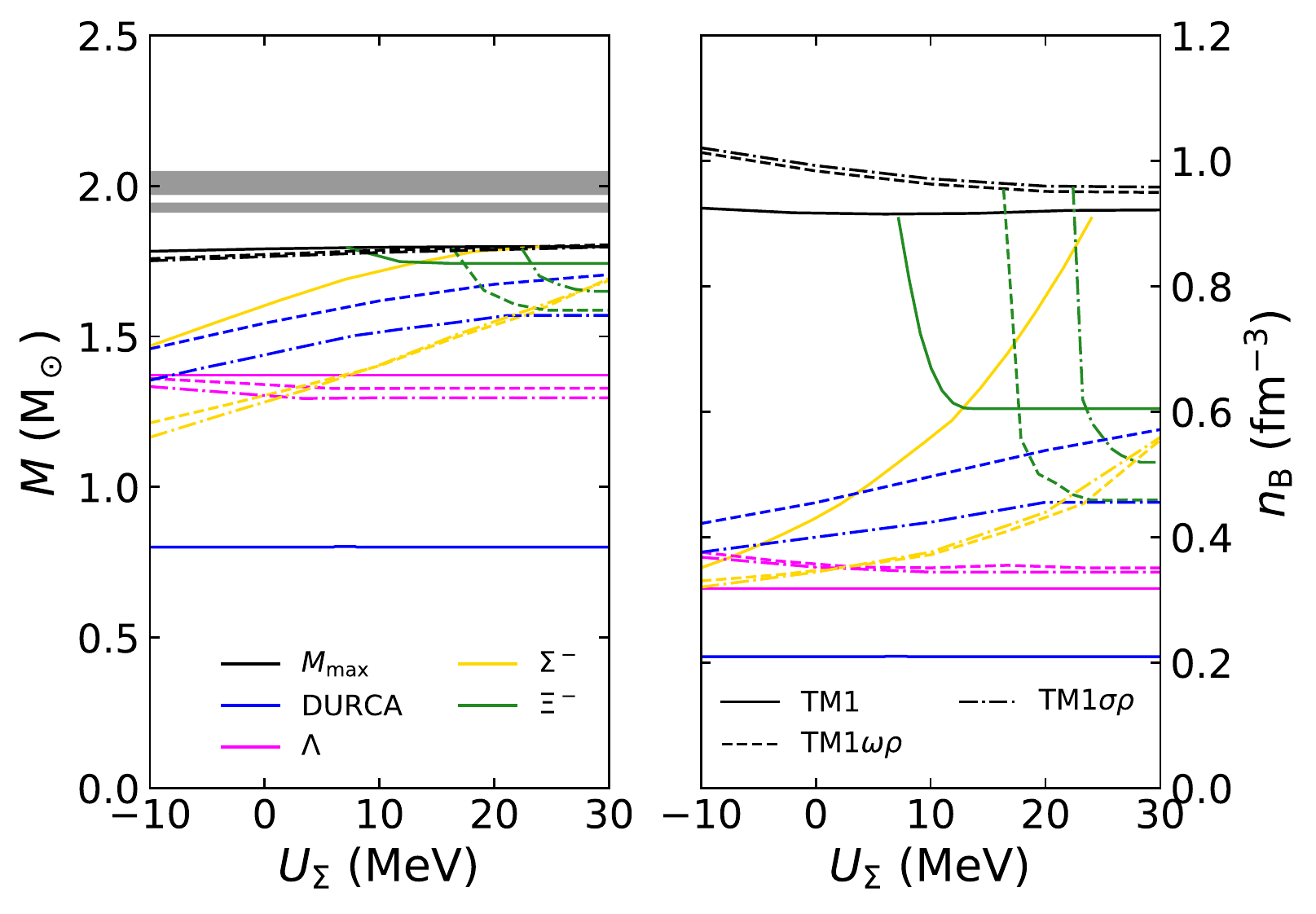}\\
\includegraphics[angle=0, width=0.45\textwidth]{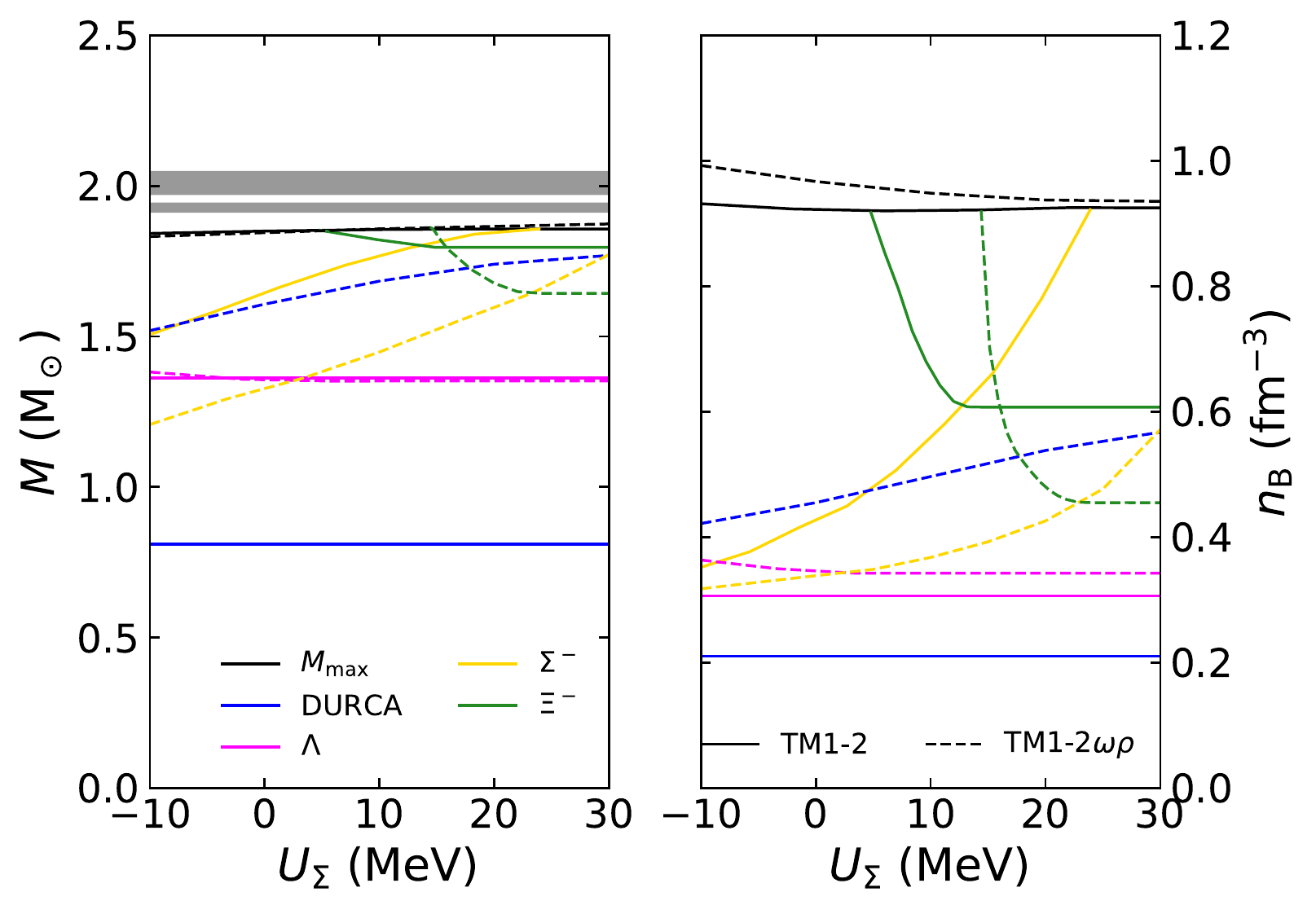}\\
 \includegraphics[angle=0, width=0.45\textwidth]{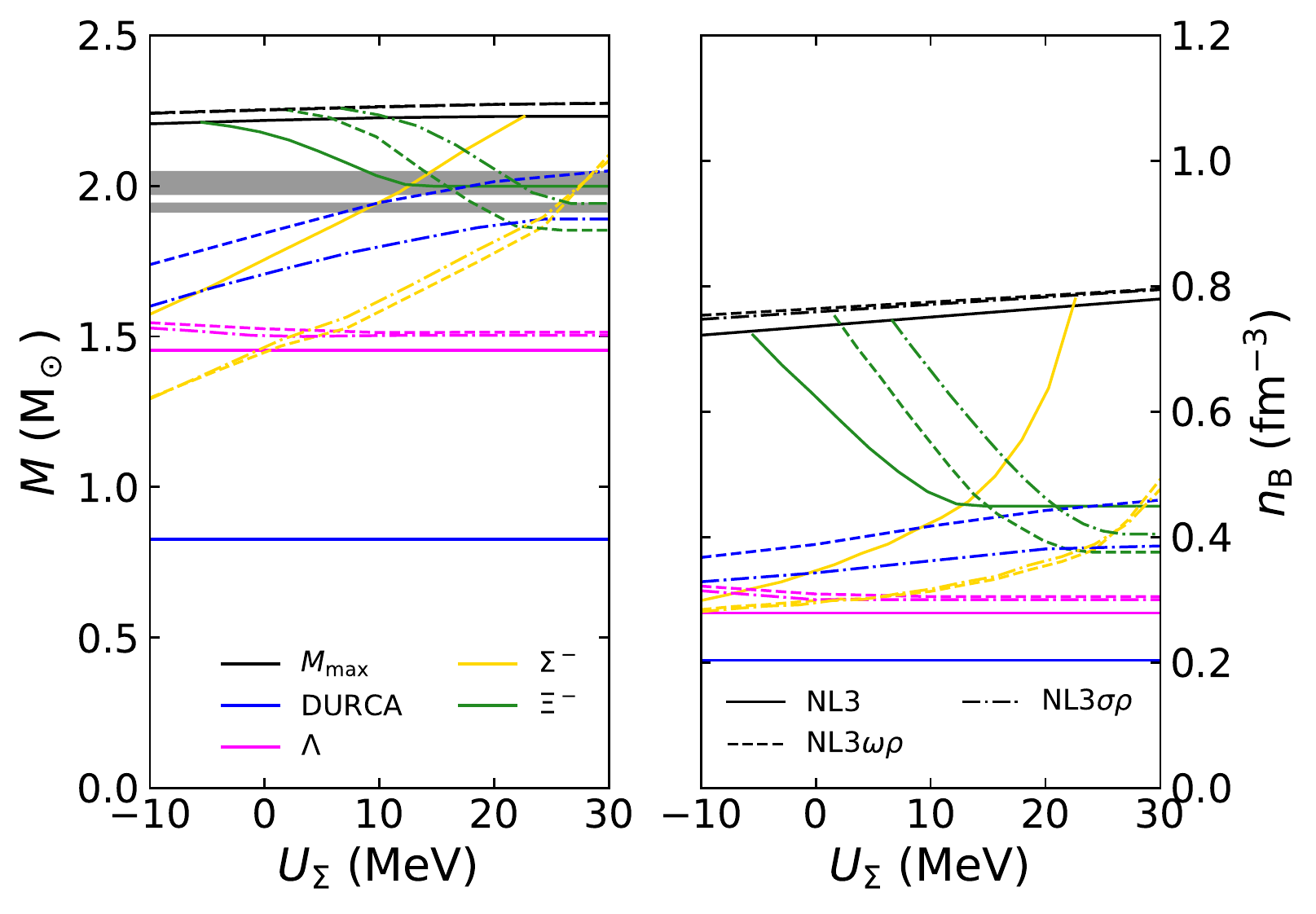}\\
\end{tabular}
\end{center}
\caption{Onset density and mass of the different hyperonsfor different values of
  the $L$ and of $U_\Sigma$ for the models
  TM1, TM1$\omega\rho$, TM1$\sigma\rho$ (top panel), TM1-2 and
  TM1-2$\omega\rho$ (middle panel) and  NL3,
   NL3$\omega\rho$ and NL3$\sigma\rho$ (bottom panel) .}
\label{tm12}
\end{figure}

In this section, we consider the set of models defined in Section
\ref{sec:model}. Most of these models have a symmetry energy slope below 60
MeV but there are three of them with a slope above 100 MeV (NL3, TM1
and FSU2), out of the
range of values  $40<L<62$ MeV \cite{Lattimer13} and $30<L<86$ MeV
\cite{Oertel17} which where defined by terrestrial, theoretical, and,
for the second range, also by observational constraints. In addition, these three models do not satisfy constraints
obtained from microscopic calculations of neutron matter based on
nuclear interactions derived from chiral effective field theory
\cite{Hebeler13}, or from realistic two- and three-nucleon interactions using
quantum Monte Carlo techniques \cite{Gandolfi12}.
 We keep them in the discussion because they are
still frequently used and it is interesting to show how a stiff
symmetry energy affects
the behavior of an hyperonic  EoS.

We have discussed in the previous section the effect of the density
dependence of the symmetry energy on the onset of the nucleonic
electron DU process, whether hyperons are included and present or not. In
Fig. \ref{sigma} left panel, we plot the DU onset density for the
different models as a function of  $U_\Sigma(n_0)$  the $\Sigma$ potential in symmetric
nuclear matter at saturation. In the right panel,
the corresponding NS masses are shown. Models with a large
$L$, i.e. NL3, TM1 and FSU2, are not affected because $n_{\rm DU}$ is just
above saturation density and lower 
than any of the hyperon onset density.  For all the other models the trend is
similar: the more repulsive $U_\Sigma(n_0)$ is, the larger
$n_{\rm DU}$.

To conclude, let us point out that the two models with density
dependent couplings do not predict the occurrence of the DU process,
even in the presence of hyperons.

\subsubsection{Hyperon species}

\begin{figure*}[t]
\begin{center}
\begin{tabular}{cc}
\includegraphics[angle=0, width=0.5\textwidth]{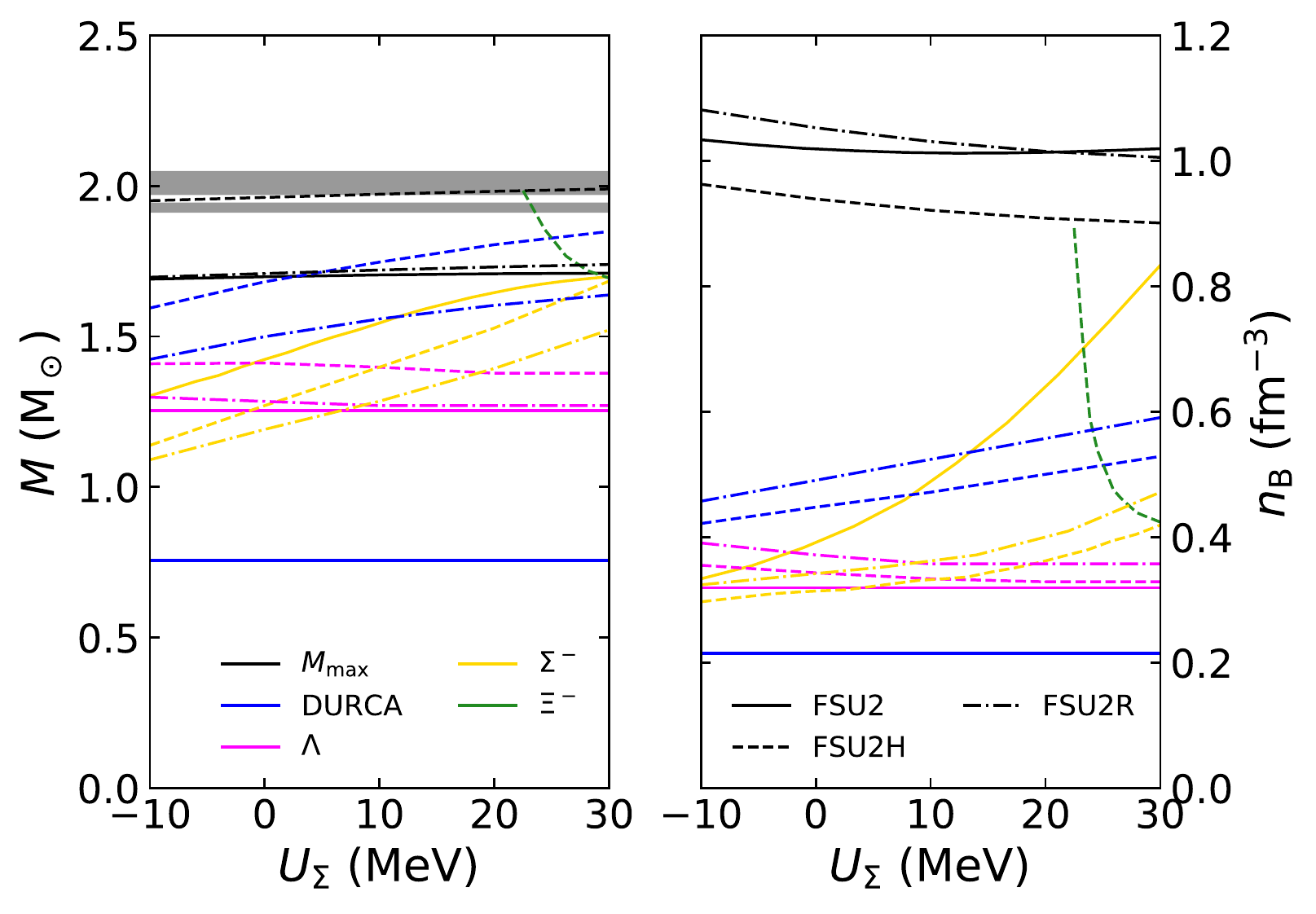}&
\includegraphics[angle=0, width=0.5\textwidth]{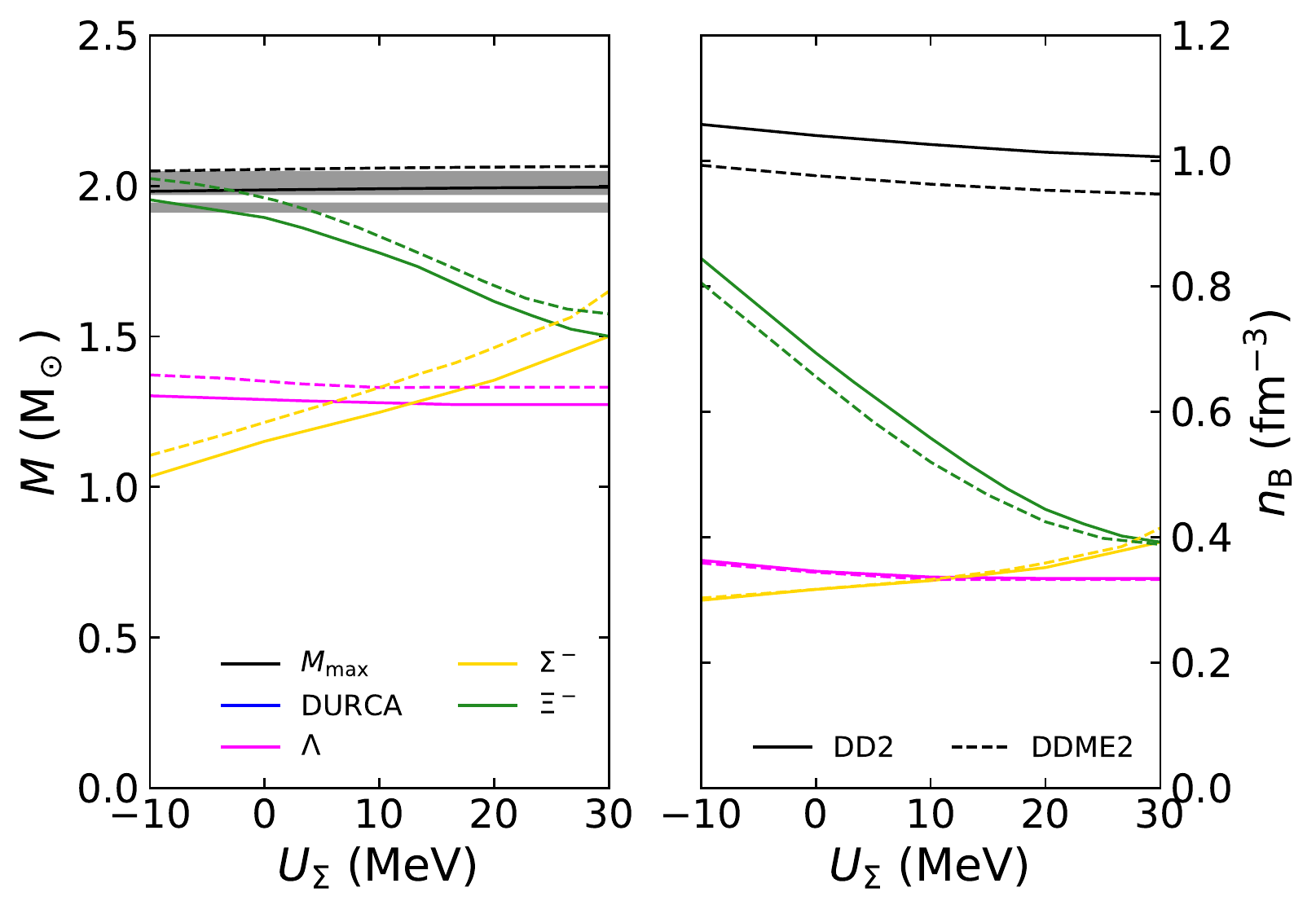}\\
\end{tabular}
\end{center}
\caption{Onset density and mass of the different hyperons for different values of
  the $L$ and of $U_\Sigma$ for the models
  FSU2, FSU2R and FSU2H (left panel) and DD2, DDME2 (right panel) .}
\label{fsu2}
\end{figure*}

In Sec. \ref{sec:hyp} we have indicated  the different channels
that allow for hyperonic direct Urca. It is, therefore,  important to
determine under which conditions these processes occur, in particular,
the masses of the NSs for which they are opened. In the present
section we show for all models of Table \ref{tab:nuclear} the maximum mass
central density, the onset
density of the
different hyperons and the onset density of the nucleonic electron DU
process as a function of the $U_\Sigma$, and the corresponding NS
masses.

In Fig. \ref{tm12} the information above is plotted for TM1,
TM1$\omega\rho$ and TM1$\sigma\rho$  MeV (top panel),
TM1-2  and TM1-2$\omega\rho$ (middle panel), and for NL3 ,
  NL3$\omega\rho$  and NL3$\sigma\rho$ (bottom panel). All the models with a nonlinear term in $\omega\rho$ or $\sigma\rho$ have $L\simeq 55$ MeV while TM1, TM1-2 and NL3 have a slope of the symmetry energy which is twice larger: $L\sim110-120$ MeV. The behavior of the TM1
and TM1-2  EoS only differ
above saturation density, the
 TM1-2 EoS  being stiffer. As a consequence, hyperons set in at lower
densities in TM1-2, and the maximum masses are larger, but still below
1.9 $M_\odot$, for the set of hyperon-meson coupling chosen which
considers for the vector-isoscalar mesons the $SU(6)$ symmetry.  For
TM1 and TM1-2 as discussed  before, the DU sets in in
NSs with masses below 1$M_\odot$, independently of $U_\Sigma$. 
Models including the nonlinear term $\omega\rho$ or $\sigma\rho$,
and having a symmetry energy slope $L\sim 55$ MeV, show a very
different behavior. In this case, the magnitude of  $U_\Sigma(n_0)$
has an important effect on the
behavior of the system: for $U_\Sigma\lesssim 5$ MeV,  the $\Sigma$
hyperon  sets in at densities below the onset of $\Lambda$, and the
corresponding NS have masses below $\sim 1.2 \, M_\odot$, that is $\sim 0.2-0.3 M_\odot$ smaller than the mass of the star where the
nucleonic electron DU process starts operating. For  $U_\Sigma\gtrsim
5$ MeV, the $\Lambda$-hyperon is the first hyperon to set in and is not
affected by the  magnitude of  $U_\Sigma(n_0)$. This occurs for
stars with a mass  $\sim 1.3 \, M_\odot$. If $U_\Sigma\gtrsim
20$ MeV, the $\Xi^-$-hyperon sets in before $\Sigma^-$, corresponding to a star mass of
$\sim 1.6 \, M_\odot$.  It is interesting  to comment  on the
differences between models TM1$\omega\rho$ and TM1$\sigma\rho$ which
have the same symmetry energy slope at saturation, but the density
dependence of the symmetry energy in  TM1$\omega\rho$ is modeled by the coupling of the
$\omega$-meson to the $\rho$-meson, while in   TM1$\sigma\rho$ the
$\rho$-meson couples to the $\sigma$-meson. Within TM1$\sigma\rho$, the
onset of the $\Lambda$ and  $\Sigma$-hyperons as well as the nucleonic
electron DU process occur in stars with lower masses. This is due to
the fact that the softening effect on the symmetry energy, which is
always very effective in   TM1$\omega\rho$ because the $\omega$-field
increases with density, saturates in model  TM1$\sigma\rho$  due to
the behavior of the $\sigma$-meson with density. Finally, we also
conclude that the overall effect of the value of  $U_\Sigma(n_0)$ on the star maximum
mass is negligible.

Similar conclusions may be drawn for the models NL3,
  NL3$\omega\rho$ and NL3$\omega\rho$, the main difference being that
  in this case much larger star masses are attained, well above  $\sim
  2 \, M_\odot$,  because these EoSs
  are harder than the EoS resulting from TM1, TM1-2 and respective
  families. For these models the maximum NS masses
  correspond to configurations where the effective nucleonic mass
  becomes zero, as already pointed out in \cite{Fortin17}.
As a consequence of the extra hardness,  the central
  densities are smaller, and for NL3$\sigma\rho$ and NL3$\omega\rho$,
  the different processes set in for
  more massive stars when compared to the TM1 like models: the $\Lambda$-hyperon appears masses at  $\sim 1.5 \,
  M_\odot$, the nucleonic electron DU process turns on above  $\sim 1.6 \,
  M_\odot$ if $U_\Sigma=-10$ MeV and $\sim 1.9 \,
  M_\odot$ if $U_\Sigma=+30$ MeV. Besides the crossing between the
  onsets of the $\Sigma$-hyperon and the $\Lambda$-hyperon occurs for
  slightly smaller values of $U_\Sigma(n_0)$ than for the TM1 models.

\begin{figure*}[t]
\begin{center}
\begin{tabular}{cc}
\includegraphics[angle=0, width=0.5\textwidth]{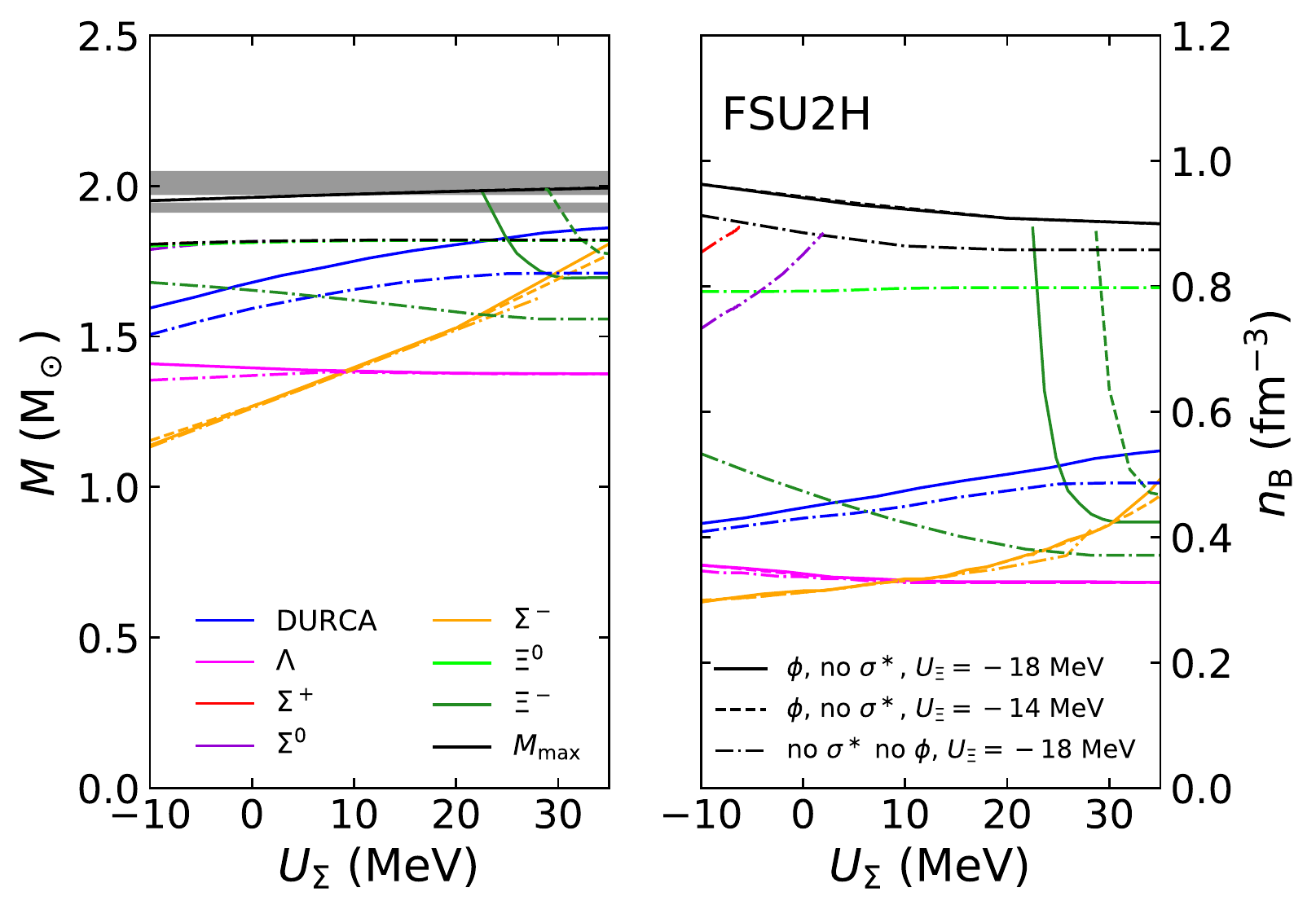}&
\includegraphics[angle=0, width=0.5\textwidth]{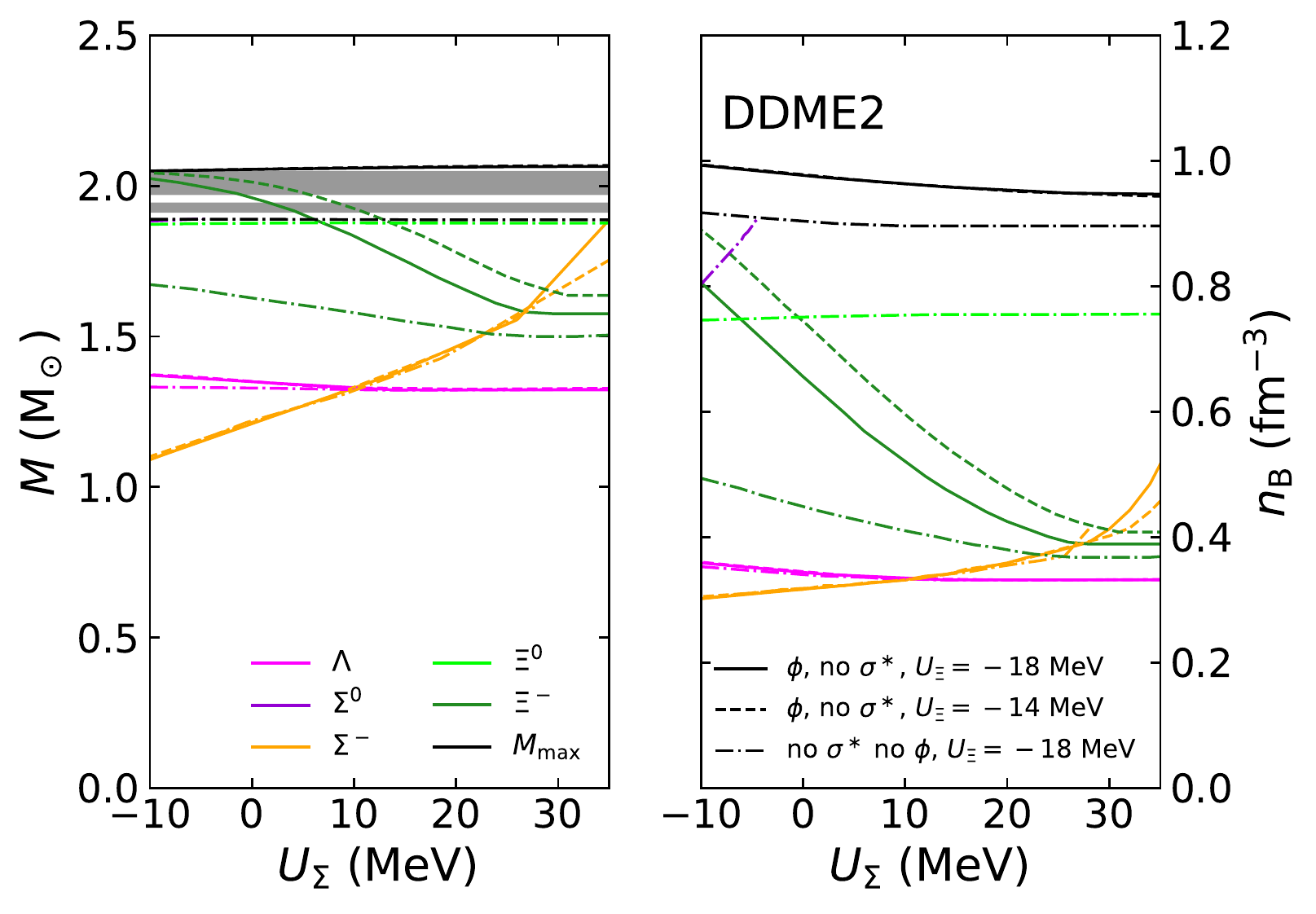}\\
\end{tabular}
\end{center}
\caption{Onset density and mass of the different hyperons for different values of
  $U_\Sigma$,  $U_\Xi$ and $x_{\phi Y}$ and  for models FSU2H and DDME2.}
\label{uxi}
\end{figure*}

In Fig. \ref{fsu2} left panel, the behavior of models FSU2, FSU2R and FSU2H is
shown. Model FSU2 has a large symmetry energy slope $L=113$ MeV, and properties
similar to the ones of TM1, presenting, however, smaller star masses
at the hyperon onset and smaller maximum star masses. FSU2R and FSU2H
have been fitted to a different set of properties and, in particular,
to a smaller symmetry energy slope ($L\sim 45$ MeV), and  were built to
describe a $2M_\odot$ star, even in the presence of
hyperons for FSU2H . FSU2 and FSU2R in fact predict similar maximum masses taking
the $SU(6)$ symmetry to fit the vector isoscalar mesons, close to 1.75
$M_\odot$, but for FSU2H the maximum mass goes up to $2M_\odot$. Comparing
the FSU2H and FSU2R models, it is clear that because FSU2H is harder,
the onset of hyperons occurs at smaller densities, which, however,
corresponds to larger star masses. As an example, the onset of $\Lambda$s
occur at $\sim 1.3M_\odot$ for FSU2R and at $\sim 1.4M_\odot$ for
FSU2H. Also the nucleonic electron DU process turns on for the FSU2H model for
masses $\sim 0.2\, M_\odot$ larger, and above $1.5  M_\odot$ whichever  values of $U_\Sigma$ is employed, going up to $\sim 1.7
M_\odot$ for $U_\Sigma=+30$ MeV. The
$\Sigma$-hyperon appears before the $\Lambda$-hyperon these
two models at larger values of $U_\Sigma$ than discussed before,
i.e. for  $U_\Sigma\lesssim +10$ MeV.  For such a slightly attractive potential hyperons appears already
in stars with masses below 1.25$M_\odot$. One
difference with respect to the previous NL3, TM1 and TM1-2-like models is that for the FSU2
like models, the $\Xi$-meson does not set in before the
$\Sigma$-hyperon for $U_\Sigma\le +30$ MeV. This is a consequence of
the large isopsin of $\Sigma^-$ that compensates the repulsion of the
$\Sigma$ potential in symmetric nuclear matter.
In order to analyze the effect of the present results on the cooling of the
NSs, one would need to take into
account the nucleonic and hyperonic pairing \cite{Raduta17,Negreiros18},
and this will be left for a future work.

We finally consider the two models with density-dependent parameters,
see Fig. \ref{fsu2} right panel. They have very similar behaviors, the only
difference being that, since the  DDME2 EoS is slightly harder, the
incompressibility at saturation is $K=251$ MeV, the onset of hyperons and of the nucleonic DU process occur at smaller densities and slightly larger
star masses ($\sim 0.1M_\odot$). Just as for the FSU2-like models, for these two
models the $\Xi^-$-hyperon does not set in before the $\Sigma^-$ for
 $U_\Sigma(n_0)$ in the range $-10,+30$ MeV. The $\Lambda $-meson appears in
stars with $M=1.3-1.4 M\odot$ and if $U_\Sigma \sim  -10$ MeV stars with
$M\sim 1- 1.1 M_\odot$ already contain $\Sigma$-hyperons. The two density-dependent models 
do not allow for the nucleonic electron DU
process to turn on. However,  the
hyperonic DU processes operate inside the stars,  and for $U_\Sigma\le
10$ MeV the process described in Eq. (\ref{y3}) is already open for stars with
$M\sim 1.3 M_\odot$.

Before finishing this section we would like to discuss the effect of the
uncertainties introduced in the previous discussion by fixing the
$U_\Xi$ in symmetric matter to $-18$ MeV and by the  unconstrained couplings
of the $\Sigma$ and $\Xi$-hyperons to the $\phi$-meson.

Following \cite{khaustov00}, we could have considered
$U_\Xi(n_0)=-14$ MeV. In Fig. \ref{uxi} the solid (dashed)  lines were
obtained with $U_\Xi=-18\, (-14)$ MeV. The curves corresponding to
these two calculations are generally superposed, except for the ones
showing the onset density of the $\Xi$-hyperon, which will occur at a
density 0.05-0.1 fm$^{-3}$ larger, if the higher value of $U_\Xi$ is
considered. All other properties, such as the onset of the DU process and of
 the other hyperons are insensitive to this change of $U_\Xi$, except if the
$\Sigma$ potential is so repulsive that the $\Xi$ hyperon sets in
before the $\Sigma$ hyperon. If future experiments show that the
$\Sigma$ potential is very repulsive in symmetric nuclear matter,
models will be more sensitive to the $\Xi$ hyperon interaction.

We discuss in the following  the role of the $\phi$ meson.  In
Fig. \ref{uxi}, for the FSU2H and DDME2 models  the 
results of switching off the coupling of the
hyperons $\Sigma$ and $\Xi$ to the $\phi$ meson (as in the minimal hyperonic models defining a lower limit on the NS mass \cite{Fortin17})
 are compared with the 
previous calculations  for which the $\phi$ couplings to $\Sigma$ and $\Xi$ hyperons are
fixed to the $SU(6)$ values.
The $\phi$ meson is responsible for the description of
the $YY$ interaction and, therefore, its effect is noticeable at high
densities but not on  the first
hyperon to appear,  for which it is the YN interaction that plays a role.  Once
the first hyperon sets in, not including the coupling to the $\phi$-meson are
results in an earlier onset (lower density) of the other hyperons. In particular,
the $\Xi$ hyperon is strongly affected because, having strangeness $-2$,
the coupling of the $\phi$ meson to the $\Xi$ hyperon is two times
larger. An immediate consequence of this last effect is that the the maximum mass configuration is lowered and for both FSU2H and
DDME2 it falls below 1.9$M_\odot$, the mass of the PSR J$1614-2230$.
Removing the $\phi$-meson also affects the DU process in the FSU2H
model, bringing its onset to lower densities, because of an increased
hyperon content and thus a reduction of the neutron Fermi momentum which ultimately favors the
occurrence of the DU process.

\subsubsection{Steady thermal state of accreting NSs}

We now explore how the value of the $U_\Sigma$ potential and of the symmetry energy affects the cooling of NSs. In particular, we model the thermal state of NSs in Soft X-ray transients (SXTs) and focus more specifially on SAX J1808.4-3658 (SAX J1808 in the following) \cite{SAX,Heinke09}, the SXT with the lowest-observed luminosity.

In SXTs NSs accrete matter from their binary companion during short phases with a high luminosity followed by long period of quiescence characterized by a low luminosity
signaling zero or strongly reduced accretion. During the accretion phases, the accreted matter undergoes a series of nuclear reactions (electron captures and pycnonuclear fusions - see \cite{HZ08} and references therein) as it sinks deeper into the crust under the weight of the newly-accreted matter. These reactions release heat in the crust which propagates in the NS interior, inwards heating the core and outwards emitted in the form of photons at the surface. This is the so-called deep crustal heating. After frequent and short periods of accretion the NS reaches a state of thermal equilibrium with a constant internal temperature throughout the star \cite{Yakovlev03,Yakovlev04}. This temperature is determined by the balance between the heating generated during the accretion phase which is directly proportional to the accretion rate $\dot{M}$ averaged over periods of accretion and quiescence, and the energy losses in the form of 1) photons emitted from the surface of the star and 2) of neutrinos freely escaping from the whole star (see  e.g. \cite{Fortin18a} for details). Consequently the steady thermal states of accreting NS depends on three ingredients 1) the composition of the NS envelope from where the photons escape; 2) the NS core properties (EoS and composition) since the core is reponsible for most of the neutrino losses; 3) the total heat release in the accreted crust. The EoS for the crust hardly affect the thermal states, only the heat release per accretd nucleon $Q_{\rm DCH}$ does and its values has been shown to be rather robust: $Q_{\rm DCH}\sim 2$ MeV per accreted nucleon \cite{HZ08,Fantina18}. Thus, in the following we adopt the model for the accreted crust and the deep crustal heating from Ref. \cite{HZ08} for lack of model consistent with the core EoSs that we employ. We use two limiting models of NS envelopes corresponding to either the absence of light elements (non-accreted envelope) or a maximum amount of them (fully accreted envelope) from Ref. \cite{Potekhin03}.

\begin{figure*}[t]
\begin{center}
\begin{tabular}{cc}
\includegraphics[angle=0, width=0.5\textwidth]{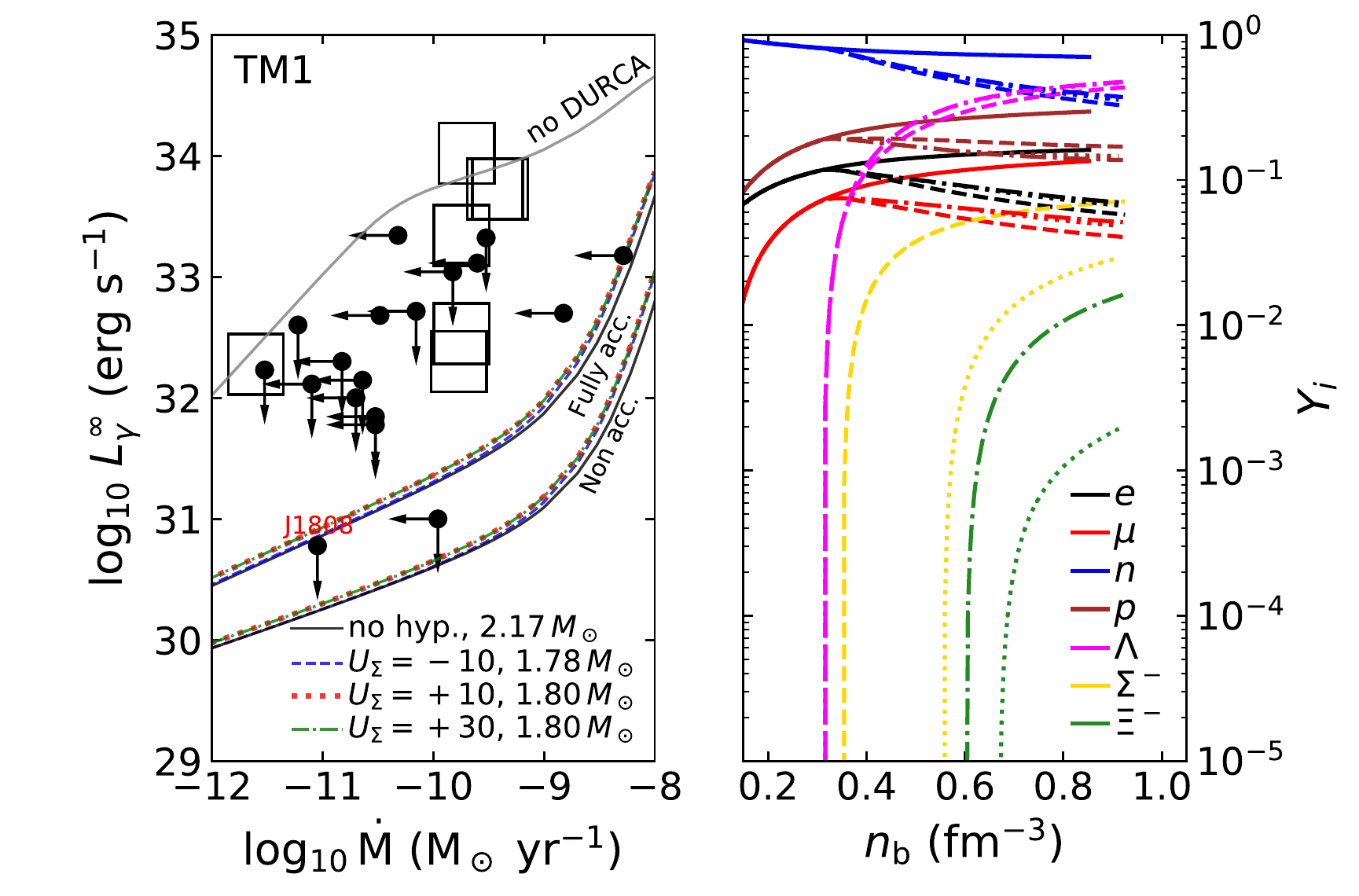}&
\includegraphics[angle=0, width=0.5\textwidth]{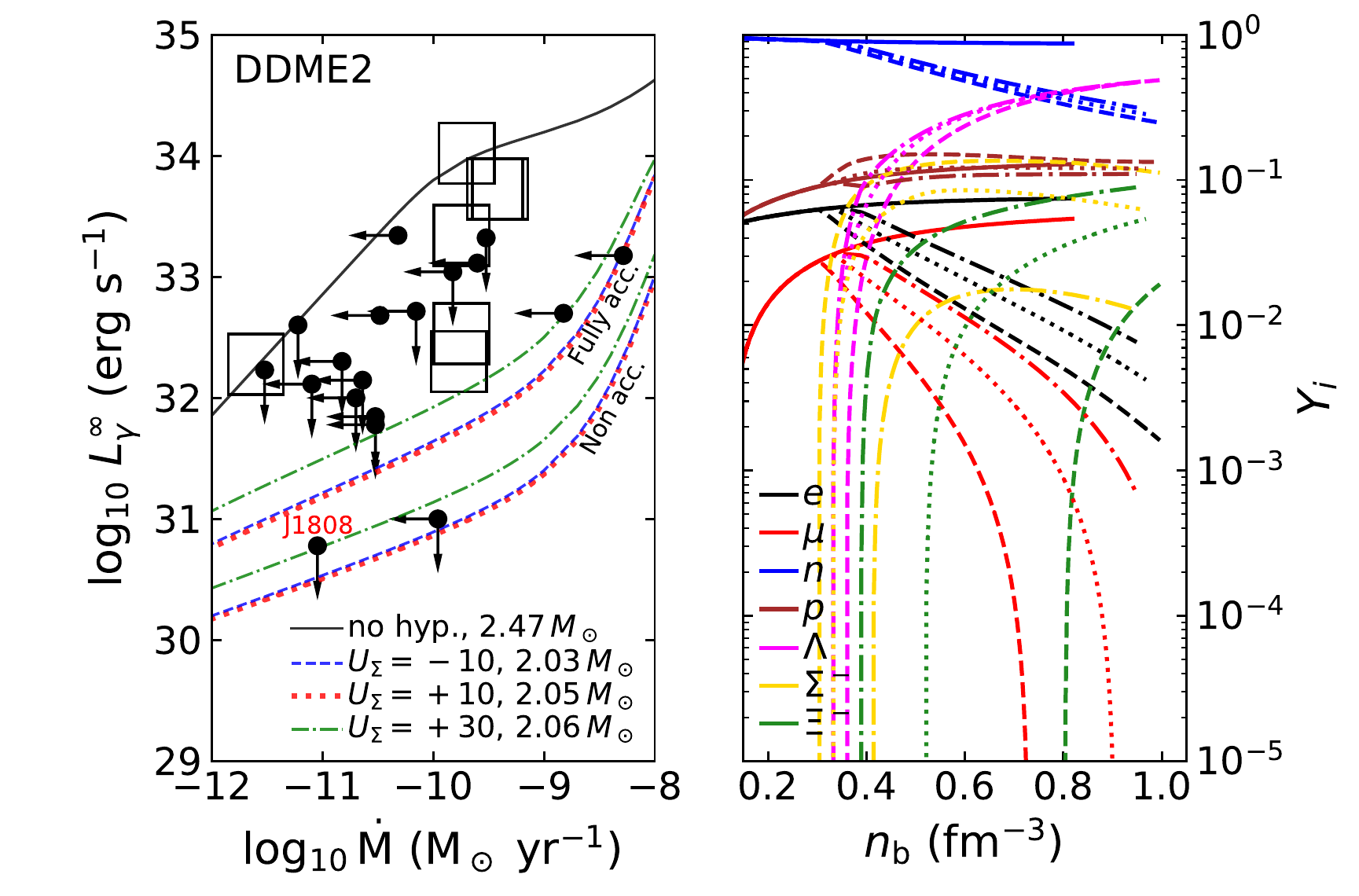}\\
\end{tabular}
\end{center}
\caption{For the TM1 (left) and DDME2 (right) parametrizations, the left panel of each plot shows the luminosity of NSs in SXTs obtained for different masses and EoSs vs the observational data taken from \cite{BY15}. In the right panel the composition of each EoS that is employed is plotted. In addition to the purely nucleonic EoS for each parametrization (solid line), hyperonic EoSs with various values of the $U_\Sigma$ potential are employed: $-10,\, 10,\, 30$ MeV (dashed, dotted and dot-dashed lines, respectively). The NS maximum mass for which the lower bound of the luminosity of SXTs is obtained are indicated in the labels of the left panels of each plot. SAX J1808, with the lowest observed luminosity, is indicated in red.}
\label{fig:cooling}
\end{figure*}

In Fig. \ref{fig:cooling} for the TM1 (left) and DDME2 (right) EoSs, we show, on the left panel of each plot the luminosity in quiescence as a function of the accretion rate together with the observational data from \cite{BY15} and on the right panel the composition for the different models. We use the TM1  and DDME2 EoSs  with various hyperonic contents
obtained for different values of the $\Sigma$ potential (dashed, dotted, and dot-dashed lines) together with their purely nucleonic versions (solid lines).
TM1 is chosen  as a representative  model that predicts that the nucleonic DU process occurs for quite
low star masses $M<0.8M_\odot$ while DDME2 as a model which does not
allow for this process at all. For each EoS we compute 1) the upper
bound on the thermal state of NSs that is obtained for NSs with a mass
below the DU threshold - this defines the lowest possible neutrino
losses and hence the largest luminosity, 2) the lower bound of the
thermal state which is reached for maximum mass NSs with the largest
neutrino emissions obtained when the DU processes operate and hence
the lowest luminosity. We do not include superfluidity in the models
(see discussion in \cite{Fortin18a}) as it reduces the DU
emissivity. We indeed want to confront the lowest-bound on the thermal
state we obtain with the observational data on SAX J1808. This object,
indicated in red in the plots in Fig.\ref{fig:cooling}, has the lowest
observed luminosity and a precisely measured accretion rate thanks to the observations of multiple type I X-ray bursts \cite{Heinke09}. Its low-luminosity is challenging to model and suggests that very efficient neutrino processes, the most efficient of which are the nucleonic and hyperonic DU processes, are operating in its NS core.  In \cite{Yakovlev04}, the authors could explain its luminosity  only by using an hyperonic  core EoS. The model they have
considered for nuclear matter is GL85 \cite{GL85} that predicts a
quite hard EoS with an incompressibility $K=285$~MeV and a symmetry
energy at saturation $E_{sym}=36.8$ MeV. For the hyperonic
interaction the universal couplings were considered, i.e. the
hyperon-meson couplings equal the nucleon-meson couplings. This choice gives rise to strongly attractive hyperon potentials in
symmetric nuclear matter at saturation, of the order $-60$ to $-70$ MeV,
and allows for the appearance of all six hyperons inside the maximum mass
star, and, therefore, all channels defined by
Eqs. (\ref{y2})-(\ref{y4}) are opened. As a consequence in addition to
the nucleonic DU process all hyperonic processes are turned on and
hence  the neutrino emissivity is larger and the luminosity lower  for
the hyperonic EoS than for purely nucleonic one. The low-luminosity of
SAX J1808 could only then be modelled for a hyperonic NS, suggesting
that hyperons could be present in SAX J1808.

For the hyperonic TM1 EoSs on the left plot in Fig. \ref{fig:cooling}, in addition to the nucleonic DU process, for the model with a slightly attractive potential, $U_\Sigma=-10$ MeV  the DU channels in Eqs. (\ref{y3}),(\ref{y1}),  (\ref{y4}) are operating in the star with the maximum mass, for a repulsive $U_\Sigma=10$ MeV the DU process in Eq. (\ref{y6}) is turned on as the $\Xi^-$ is present. However since the $\Sigma^-$ appears at larger densities than when an attractive potential is used, the most efficient of all hyperonic DU processes  turned on for such models, is the one in Eq. (\ref{y3}) that then operates in a smaller region of the star and the process in  Eq. (\ref{y6}) is too weak to compensate these lesser neutrinos losses. For the model with $U_\Sigma=30$ MeV since no $\Sigma^-$ are present only processes in Eqs. (\ref{y1}) and (\ref{y6}) set in and both are less efficient than the one in  Eq. (\ref{y6}). Hence the model with the $U_\Sigma=-10$ MeV  is the coolest of all hyperonic models.
We obtain that the purely nucleonic has the lowest luminosity compared
to hyperonic models but the difference is quite small. The purely
nucleonic NS, in which only the nucleonic DU process, which is the
most efficient process, operates  is almost $\sim0.2M_\odot$ more
massive than the hyperonic NSs. Hence  for  hyperonic NSs even if more
DU channels are opened,  these are less efficient and do not exactly
compensate for the fact that the nucleonic NS has an extra region of $0.2M_\odot$ emitting neutrinos via the most efficient channel. Thus hyperonic stars emit all in all less neutrinos and hence have a slightly larger luminosity. As in \cite{Fortin18a} we obtain that NSs with a fully accreted envelope are more luminous than with a non-accreted one. Thus we obtain that for the TM1 EoS SAX J1808 is compatible with a NS with a small or null amount of accreted matter in the envelope, with or without hyperons.

For the DD2 parametrization (right plots of Fig. \ref{fig:cooling}), as the nucleonic DU process does not operate at all for the purely nucleonic EoS, non-hyperonic NSs will have a very similar and large luminosity. Hyperonic models have, however, a small luminosity as the additional hyperonic DU processes operates and only such models can explain the low-luminosity of SAX J1808. For all hyperonic models the $\Xi^-$ , $\Sigma^-$ and $\Lambda$ are present at the maximum mass, and the latter two  species in similar amount. The most efficient hyperonic DU process is then the channel in Eq. (\ref{y3}) between the $\Lambda$ and the $\Sigma^-$. As the model with $U_\Sigma=-10$ MeV has the largest amount of $\Sigma^-$ (it even appears before the $\Lambda$) it has the largest neutrino emissivity and hence the lowest luminosity of all models. The model with  $U_\Sigma=10$ MeV has approximately 50 \% less of  $\Sigma^-$ and hence is slightly more luminous as it emits less neutrinos. Finally for $U_\Sigma=30$ MeV the fraction of $\Sigma^-$ is one order of magnitude less than for the slightly attractive potential. As a consequence this model gives the largest luminosity of all hyperonic models. We conclude  that for the DDME2 model, since the nucleonic DU process does not operate, SAX J1808 is only compatible with a NS with hyperons and no or a very small amount of accreted matter in the envelope.

We can see that the delicate interplay between the symmetry energy and the $\Sigma$-potential strongly affects the cooling of SXTs. These objects could potentially offer the possibility to constraint the  $\Sigma$-potential and thus the properties of the $\Sigma$ hyperon, from the astrophysical observations of SXTs with a low-luminosity complementing the little experimental constraints on the properties of the $\Sigma$ hyperon currently available. A more systematic study of the thermal state of accreting NSs is beyond the scope of the present paper and will be the subject of a future work.

\subsubsection{Hyperonic star radius}

\begin{figure*}[t]
\begin{center}
\begin{tabular}{c}
\includegraphics[angle=0, width=0.9\textwidth]{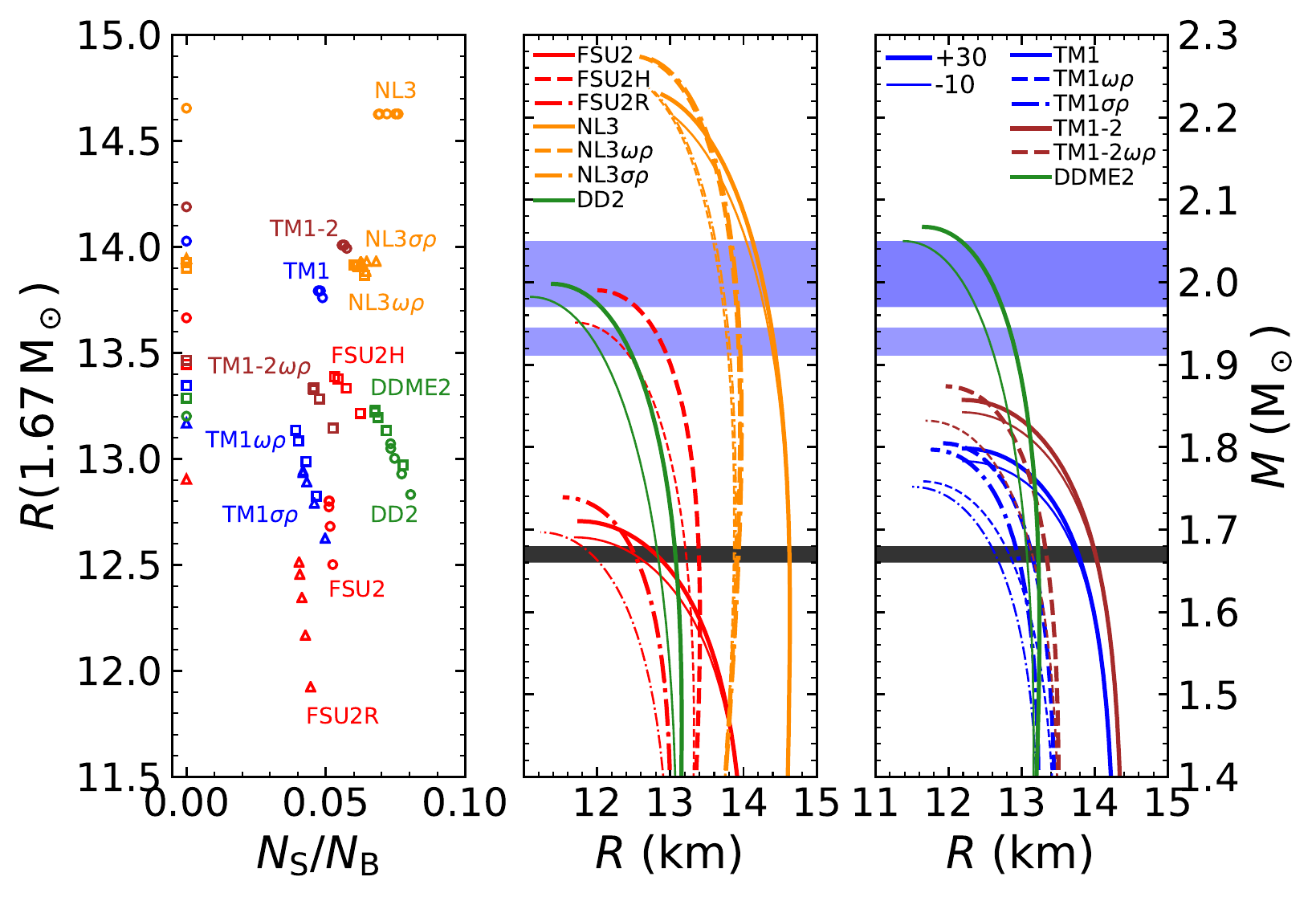}\\
\end{tabular}
\end{center}
\caption{\it 
Left: radius of a $1.67M_\odot$ star as a function of the total strangeness fraction for all the models considered in this work and for $U_\Sigma=-10,0,10,20$ and 30 MeV. Middle and right: $M-R$ relations for all the models for $U_\Sigma=-10$ and 30 MeV (thick and thin lines, respectively.}
\label{rns}
\end{figure*}

There are still large observational uncertainties associated with the
radius of NSs including the  canonical NS with a mass equal to $1.4 M_\odot$,
see the discussion in \cite{Potekhin14,Fortin15,Steiner16},
although there have been several indirect predictions from different
analysis. 
Recently several studies have used the detection  of the gravitational
waves emitted from a neutron star merger GW170817
\cite{LIGO17} to constrain the upper limit of the $1.4
M_\odot$ star  radius to $\sim 13.7$ km
\cite{Abbott18,Annala18,Fattoyev18,Most18,Raithel18,Tews18,Malik18,Lim18}. Similar
constraints had been obtained before  from the analysis of the experimental
constraints set on the symmetry energy \cite{Li06,Steiner16}. 

Since we
are interested in analysing the effect of strangeness on the radius of
a NS, and as we have seen   for many  models,
strangeness sets in inside stars with a mass above $1.4
M_\odot$, we will consider a more massive star. In the discussion of
this section we calculate the radius of a star with $M=1.67
M_\odot$, the mass of the pulsar PSR J1903+0327. 
Results are plotted in  Fig. \ref{rns} left panel as a function of the
total star
hyperon fraction. 
 On the right panel, we
have plotted the hyperonic star mass-radius curves to help the
discussion. The thin (thick) lines correspond to $U_\Sigma=-10$ (+30)
MeV.

The strangeness fraction increases if the $\Sigma$ potential
becomes less repulsive, and simultaneously the radius decreases. The
relation between the radius and  the strangeness fraction is essentially
linear but the slope is model dependent. For models like NL3, TM1,
TM1-2 changing $U_\Sigma$ does not have a large effect on the
strangeness content and on the radius. This is clearly understood
looking at Fig. \ref{tm12} where the star mass at the onset of the
$\Sigma$ hyperon is plotted:  a star with $M=1.67M_\odot$ has no (only
a few) 
$\Sigma$ hyperons for $U_\Sigma=+30$ (-10) MeV.  Density-dependent
models have a similar behavior, being the models that predict a larger
amount of strangeness, as large as 0.075, although still satisfying the $2 M_\odot$
constraint. For $-10<U_\Sigma< +30$ MeV the radius increases $\sim300$
m. Models TM1$\omega\rho$,  TM1$\sigma\rho$,  TM1-2$\omega\rho$, FSU2H have a
similar behavior but do not predict strangeness contents above
0.05. Models FSU2 and FSU2R suffer a quite large radius change for a
small increase of strangeness because, as seen in the right panel,
1.67$M_\odot$ is very close to the maximum star mass. Contrary to
\cite{Providencia13} we do not see  a linear correlation if also the
$N_S/N_B=0$ radius is included. In \cite{Providencia13} $N_S/N_B$ is the
strangeness fraction and not the hyperon fraction. However,  in that work the
authors did not use unified crust-core EoS and different hyperon
interactions, giving rise to much larger strangeness fractions inside
the star, were discussed.

\section{Summary and conclusions}
\label{sec:concl}

In the present study, we have explored how the density dependence of
the symmetry energy may affect the properties of hyperonic neutron
stars. The study was undertaken within the RMF approach to nuclear
matter and models that describe ground-state  properties  of nuclei and
$\Lambda$-hypernuclei, as well as constraints from microscopic
calculations of NS (except for three models) and the $2M_\odot$ constraint on nucleonic stars have been
chosen.  We have also considered a family of models based on TM1
\cite{tm1,Providencia13,Bao2014} that has allowed us to directly discuss the
effect of the density dependence of the symmetry energy on the
properties of hyperonic stars. For all the models considered, we have
taken an inner crust-core unified EoS. In the present work, we have
calculated the FSU2, FSUR2H and FSU2H inner crust of catalyzed
$\beta$-equilibrium matter, which are given as Suplementary Material.

 The $\Lambda$-meson and $\Xi$-meson couplings   were constrained  by the existing
hypernuclei experimental data. Taking into account  the present lack of
knowledge concerning the properties of the $\Sigma$ hyperon in nuclear
matter, we have discussed the properties of hyperonic matter
considering values of the $\Sigma$ potential in symmetric nuclear
matter that go from $-10$ MeV to +30 MeV at saturation density, having
in mind that if no $\Sigma$-hypernucleus has been detected, the
$\Sigma$ potential must be repulsive or only slightly attractive.

We have shown  that  the  DU process
is affected by hyperons only if the slope of the symmetry energy is
$L\lesssim 70$ MeV.
The nucleonic electron DU process is both sensitive to the slope of
the symmetry energy and, for $L\lesssim 70$ MeV, to
the value of the $\Sigma$ potential in nuclear matter.  The more
repulsive $U_\Sigma$ the larger the  nucleonic electron DU process.
A small $L$
shifts the DU onset to larger densities but the effect is stronger 
the more repulsive the $\Sigma$ potential is. Models with density-dependent couplings  simply 
do not allow for the nucleonic
electron DU process to turn on. However, the cooling of  stars within this framework is also affected when new hyperonic channels open
inside the star. So, even though the density-dependent models do not predict
nucleonic electron DU, when the reactions  described in Eqs.  (\ref{y1}),
(\ref{y3}) and   (\ref{y4}) start to operate the star
is much less luminous.  This occurs in stars with a mass of
the order of $1.1 -1.3 M_\odot$ models.
All other models, with constant couplings, predict the occurrence of both hyperonic and nucleonic
DU processes inside massive enough NSs. 

We have studied  how the value of the $U_\Sigma$ potential affects the
thermal state of NSs in Soft X-ray transients and  focused more
specifially on SAX J1808 \cite{SAX,Heinke09}, the SXT with the
lowest-observed luminosity. We have shown that the low luminosity of
this object could be described by a model,  with an unrealistically
high symmetry energy slope, that predicts the opening of
the DU inside  low mass stars, independently of taking a nucleonic or
an hyperonic EoS. For the nucleonic EoS, the maximum star mass is
large and allows the nucleonic DU process to occur in a wide range of the NS
interior, while for the hyperonic EoS although the maximum mass is smaller, inside the
 core both the nucleonic DU and the hyperonic DU processes
act. 
However, the SAX  J1808  low luminosity could also be explained in
 the framwork of a density dependent hadronic model, satisifying
well established nuclear matter and nuclei properties and describing a
2$M_\odot$ star, if hyperonic degrees of freedom are
allowed to occur inside the star.  In this case, objects like the
SAX J1808 could potentially offer the possibility to constraint the
hyperonic interaction, in particular, the $\Sigma$ potential.

{\bf Acknowledgments}:
This  work  was  supported  by  Fundação  para  a
Ciência e Tecnologia,  Portugal,  under the projects
UID/FIS/04564/2016   and   POCI-01-0145-FEDER-
029912 with financial support from POCI, in its FEDER
component, and by the FCT/MCTES budget through
national  funds  (OE), and by the Polish National Science Centre (NCN) under
759 grant No. UMO-2014/13/B/ST9/02621.  Partial  support  comes  also from
PHAROS COST Action CA16214.  H.P.
is supported by FCT (Portugal) under Project No.
SFRH/BPD/95566/2013


%


\section{Appendix}
\begin{table*}[htb]
\label{eosc}
\caption{ Equation of state of the inner crust with pasta for the FSU2, FSU2R, and FSU2H models. The energy density, $\varepsilon$, and pressure, $P$,  are in units of fm$^{-4}$. }  \label{tab1}
  \begin{tabular}{c|ccccccccccccccccccc}
\hline\hline
             &   \multicolumn{2}{c}{FSU2} &\phantom{a} &\multicolumn{2}{c}{FSU2R} &\phantom{a} &\multicolumn{2}{c}{FSU2H} &\phantom{a} \\
$n_B$ (fm$^{-3}$)  &  \multicolumn{1}{c}{$\varepsilon$} & \multicolumn{1}{c}{P}  &\phantom{a} &  \multicolumn{1}{c}{$\varepsilon$} & \multicolumn{1}{c}{P} &\phantom{a} &  \multicolumn{1}{c}{$\varepsilon$} & \multicolumn{1}{c}{P}   \\
\hline
0.002	&	-				&	-				&\phantom{a}	&	0.009527397342	&	1.114900624E-05	&\phantom{a}	&	0.009527062997	&	1.084494306E-05	\\
0.003	&	0.01427514106	&	1.190916646E-05	&\phantom{a}	&	0.014298330992	&	1.92066982E-05	&\phantom{a}	&	0.014297628775	&	1.854789298E-05	\\
0.004	&	0.019038049504	&	1.555793278E-05	&\phantom{a}	&	0.019072251394	&	2.883538582E-05	&\phantom{a}	&	0.0190710444	&	2.77711606E-05	\\
0.005	&	0.023801861331	&	1.90039882E-05	&\phantom{a}	&	0.0238487795	&	3.988303797E-05	&\phantom{a}	&	0.023846937343	&	3.831203867E-05	\\
0.006	&	0.02856634371	&	2.219665839E-05	&\phantom{a}	&	0.028627665713	&	5.224829874E-05	&\phantom{a}	&	0.028625067323	&	5.00691749E-05	\\
0.007	&	0.033331338316	&	2.518662041E-05	&\phantom{a}	&	0.033408716321	&	6.588049291E-05	&\phantom{a}	&	0.033405266702	&	6.304256385E-05	\\
0.008	&	0.038096740842	&	2.787251651E-05	&\phantom{a}	&	0.038191791624	&	8.06275857E-05	&\phantom{a}	&	0.038187392056	&	7.702950097E-05	\\
0.009	&	0.042862471193	&	3.045706035E-05	&\phantom{a}	&	0.042976766825	&	9.633754962E-05	&\phantom{a}	&	0.042971335351	&	9.21820174E-05	\\
0.01	&	0.047628492117	&	3.299092714E-05	&\phantom{a}	&	0.047763541341	&	0.0001129090306	&\phantom{a}	&	0.047757018358	&	0.0001081453593	\\
0.011	&	0.052394766361	&	3.547411325E-05	&\phantom{a}	&	0.052552033216	&	0.0001304940524	&\phantom{a}	&	0.052544362843	&	0.000125274295	\\
0.012	&	0.057161271572	&	3.816001117E-05	&\phantom{a}	&	0.057342153043	&	0.0001487378759	&\phantom{a}	&	0.057333290577	&	0.0001430620323	\\
0.013	&	0.061928000301	&	4.089658614E-05	&\phantom{a}	&	0.062133830041	&	0.0001677418768	&\phantom{a}	&	0.062123749405	&	0.0001618632959	\\
0.014	&	0.066694952548	&	4.403857383E-05	&\phantom{a}	&	0.06692700088	&	0.0001874553564	&\phantom{a}	&	0.066915675998	&	0.0001813740673	\\
0.015	&	0.071462139487	&	4.753530811E-05	&\phantom{a}	&	0.071721583605	&	0.0002076755918	&\phantom{a}	&	0.071709007025	&	0.0002014929632	\\
0.016	&	0.076229587197	&	5.148813943E-05	&\phantom{a}	&	0.076517544687	&	0.0002283012436	&\phantom{a}	&	0.07650372386	&	0.0002223720076	\\
0.017	&	0.080997288227	&	5.609977597E-05	&\phantom{a}	&	0.08131480962	&	0.0002493830107	&\phantom{a}	&	0.081299744546	&	0.0002436057839	\\
0.018	&	0.08576527983	&	6.131953705E-05	&\phantom{a}	&	0.0861133039	&	0.0002707181557	&\phantom{a}	&	0.086097031832	&	0.0002654477139	\\
0.019	&	0.090533591807	&	6.740081153E-05	&\phantom{a}	&	0.090913005173	&	0.0002924080472	&\phantom{a}	&	0.090895555913	&	0.0002876950603	\\
0.02	&	0.09530223906	&	7.4292926E-05	&\phantom{a}	&	0.095713868737	&	0.0003139965702	&\phantom{a}	&	0.095695272088	&	0.0003102971241	\\
0.021	&	0.100071251392	&	8.219858137E-05	&\phantom{a}	&	0.100515827537	&	0.0003359398397	&\phantom{a}	&	0.100496120751	&	0.0003333046334	\\
0.022	&	0.104840673506	&	9.121913899E-05	&\phantom{a}	&	0.105318851769	&	0.0003578324395	&\phantom{a}	&	0.105298064649	&	0.0003565148218	\\
0.023	&	0.109610520303	&	0.000101202575	&\phantom{a}	&	0.11012288928	&	0.0003798264079	&\phantom{a}	&	0.110101081431	&	0.0003796743404	\\
0.024	&	0.114380836487	&	0.0001125542913	&\phantom{a}	&	0.114927917719	&	0.0004017696483	&\phantom{a}	&	0.114905133843	&	0.0004031886056	\\
0.025	&	0.119151651859	&	0.0001250715868	&\phantom{a}	&	0.119733855128	&	0.0004236115783	&\phantom{a}	&	0.119710162282	&	0.0004268042394	\\
0.026	&	0.123922996223	&	0.0001390585239	&\phantom{a}	&	0.124540701509	&	0.0004456055467	&\phantom{a}	&	0.124516174197	&	0.0004503185046	\\
0.027	&	0.12869489193	&	0.0001543630642	&\phantom{a}	&	0.129348397255	&	0.0004673461081	&\phantom{a}	&	0.129323080182	&	0.0004739848082	\\
0.028	&	0.133467406034	&	0.0001713399688	&\phantom{a}	&	0.134156942368	&	0.000488934631	&\phantom{a}	&	0.13413092494	&	0.0004976511118	\\
0.029	&	0.138240531087	&	0.0001899385388	&\phantom{a}	&	0.138966232538	&	0.0005105231539	&\phantom{a}	&	0.138939589262	&	0.0005212160759	\\
0.03	&	0.143014326692	&	0.0002100067359	&\phantom{a}	&	0.143776282668	&	0.0005318076001	&\phantom{a}	&	0.143749088049	&	0.0005448316806	\\
0.031	&	0.147788822651	&	0.00023195002	&\phantom{a}	&	0.148587062955	&	0.0005528387264	&\phantom{a}	&	0.148559391499	&	0.0005683459458	\\
0.032	&	0.152564063668	&	0.0002554136154	&\phantom{a}	&	0.153398528695	&	0.0005738697946	&\phantom{a}	&	0.153370469809	&	0.0005917081726	\\
0.033	&	0.157340064645	&	0.0002810056321	&\phantom{a}	&	0.158210650086	&	0.0005944954464	&\phantom{a}	&	0.158182263374	&	0.0006148677203	\\
0.034	&	0.162116870284	&	0.0003082193434	&\phantom{a}	&	0.163023427129	&	0.0006149184192	&\phantom{a}	&	0.162994787097	&	0.0006379765691	\\
0.035	&	0.166894495487	&	0.000337612204	&\phantom{a}	&	0.167836785316	&	0.0006351386546	&\phantom{a}	&	0.167807996273	&	0.0006608827389	\\
0.036	&	0.171672984958	&	0.0003686773998	&\phantom{a}	&	0.172650724649	&	0.0006551561528	&\phantom{a}	&	0.1726218611	&	0.0006836875109	\\
0.037	&	0.176452368498	&	0.000402124424	&\phantom{a}	&	0.177465245128	&	0.0006747682928	&\phantom{a}	&	0.177436366677	&	0.0007060868666	\\
0.038	&	0.181232705712	&	0.00043759853	&\phantom{a}	&	0.182280123234	&	0.0006842956063	&\phantom{a}	&	0.182251513004	&	0.0007283848827	\\
0.039	&	0.1860139817	&	0.0004748970096	&\phantom{a}	&	0.187095478177	&	0.0007031476125	&\phantom{a}	&	0.187067225575	&	0.0007504802197	\\
0.04	&	0.190796226263	&	0.0005146786571	&\phantom{a}	&	0.191911309958	&	0.0007218982209	&\phantom{a}	&	0.191883504391	&	0.0007723727613	\\
0.041	&	0.195579528809	&	0.0005562847364	&\phantom{a}	&	0.196727633476	&	0.0007402940537	&\phantom{a}	&	0.196700364351	&	0.0007939613424	\\
0.042	&	0.200363859534	&	0.0006004753523	&\phantom{a}	&	0.201544389129	&	0.0007583858096	&\phantom{a}	&	0.201517611742	&	0.0008047556039	\\
0.043	&	0.205149263144	&	0.0006465410697	&\phantom{a}	&	0.206361606717	&	0.000776224304	&\phantom{a}	&	0.206335306168	&	0.0008256346337	\\
0.044	&	0.209935769439	&	0.0006952419062	&\phantom{a}	&	0.211179211736	&	0.0007938092458	&\phantom{a}	&	0.21115347743	&	0.000846361625	\\
0.045	&	0.214723423123	&	0.0007457671454	&\phantom{a}	&	0.215997248888	&	0.0008110902854	&\phantom{a}	&	0.215972140431	&	0.0008667845977	\\
0.046	&	0.219512179494	&	0.0007990797167	&\phantom{a}	&	0.220815643668	&	0.0008282191702	&\phantom{a}	&	0.220791265368	&	0.0008869034355	\\
0.047	&	0.224302142859	&	0.0008542166324	&\phantom{a}	&	0.22563444078	&	0.000844891998	&\phantom{a}	&	0.225610807538	&	0.0009068703512	\\
0.048	&	0.229093328118	&	0.000912090065	&\phantom{a}	&	0.230453595519	&	0.0008614128456	&\phantom{a}	&	0.230430826545	&	0.0009265838307	\\
0.049	&	0.233885720372	&	0.0009717879584	&\phantom{a}	&	0.235273063183	&	0.0008776801988	&\phantom{a}	&	0.235251218081	&	0.0009459425928	\\
0.05	&	0.238679364324	&	0.001034171786	&\phantom{a}	&	0.240092903376	&	0.0008935928927	&\phantom{a}	&	0.24007204175	&	0.0009650985594	\\

    \hline
    \hline
  \end{tabular}
\end{table*}

\begin{table*}[htb]
\caption{ \textit{(Continued.)} }  \label{tab2}
  \begin{tabular}{c|ccccccccccccccccccc}
\hline\hline
             &   \multicolumn{2}{c}{FSU2} &\phantom{a} &\multicolumn{2}{c}{FSU2R} &\phantom{a} &\multicolumn{2}{c}{FSU2H} &\phantom{a} \\
$n_B$ (fm$^{-3}$)  &  \multicolumn{1}{c}{$\varepsilon$} & \multicolumn{1}{c}{P}  &\phantom{a} &  \multicolumn{1}{c}{$\varepsilon$} & \multicolumn{1}{c}{P} &\phantom{a} &  \multicolumn{1}{c}{$\varepsilon$} & \multicolumn{1}{c}{P}   \\
\hline
0.051	&	0.243474245071	&	0.001098278561	&\phantom{a}	&	0.244913056493	&	0.0009093027911	&\phantom{a}	&	0.244893237948	&	0.0009840012062	\\
0.052	&	0.24827042222	&	0.001162892091	&\phantom{a}	&	0.249733552337	&	0.0009247594862	&\phantom{a}	&	0.249714821577	&	0.001002650475	\\
0.053	&	0.25306776166	&	0.001220613485	&\phantom{a}	&	0.2545543015	&	0.0009399626288	&\phantom{a}	&	0.254536747932	&	0.001021046308	\\
0.054	&	0.257866412401	&	0.001290649641	&\phantom{a}	&	0.259375363588	&	0.0009549123934	&\phantom{a}	&	0.259359031916	&	0.001039239462	\\
0.055	&	-				&	-				&\phantom{a}	&	0.2641967237	&	0.0009696595371	&\phantom{a}	&	0.264181643724	&	0.001057128538	\\
0.056	&	-				&	-				&\phantom{a}	&	0.269018322229	&	0.0009840518469	&\phantom{a}	&	0.269004613161	&	0.001074814936	\\
0.057	&	-				&	-				&\phantom{a}	&	0.273840218782	&	0.0009982922347	&\phantom{a}	&	0.273827910423	&	0.001092247898	\\
0.058	&	-				&	-				&\phantom{a}	&	0.278662353754	&	0.001012228429	&\phantom{a}	&	0.278651505709	&	0.00110942754	\\
0.059	&	-				&	-				&\phantom{a}	&	0.283484727144	&	0.001025962061	&\phantom{a}	&	0.283475399017	&	0.001126353745	\\
0.06	&	-				&	-				&\phantom{a}	&	0.288307338953	&	0.001039492781	&\phantom{a}	&	0.288299590349	&	0.001143077272	\\
0.061	&	-				&	-				&\phantom{a}	&	0.293129920959	&	0.001004930935	&\phantom{a}	&	0.293124079704	&	0.001159547362	\\
0.062	&	-				&	-				&\phantom{a}	&	0.297952502966	&	0.001019424642	&\phantom{a}	&	0.29794883728	&	0.001175764133	\\
0.063	&	-				&	-				&\phantom{a}	&	0.302775323391	&	0.001033664914	&\phantom{a}	&	0.302773833275	&	0.001191778225	\\
0.064	&	-				&	-				&\phantom{a}	&	0.307598352432	&	0.001047702623	&\phantom{a}	&	0.307599157095	&	0.001207640162	\\
0.065	&	-				&	-				&\phantom{a}	&	0.312421619892	&	0.001061537536	&\phantom{a}	&	0.312424719334	&	0.001223147381	\\
0.066	&	-				&	-				&\phantom{a}	&	0.317245006561	&	0.001075220411	&\phantom{a}	&	0.317250490189	&	0.001238451921	\\
0.067	&	-				&	-				&\phantom{a}	&	0.322068631649	&	0.001088599092	&\phantom{a}	&	0.322076499462	&	0.001253503142	\\
0.068	&	-				&	-				&\phantom{a}	&	0.326892495155	&	0.001101825968	&\phantom{a}	&	0.326902478933	&	0.00121645804	\\
0.069	&	-				&	-				&\phantom{a}	&	0.331716567278	&	0.001114748651	&\phantom{a}	&	0.331728547812	&	0.001232573413	\\
0.07	&	-				&	-				&\phantom{a}	&	0.33654075861	&	0.001127468655	&\phantom{a}	&	0.336554706097	&	0.00124843535	\\
0.071	&	-				&	-				&\phantom{a}	&	0.34136518836	&	0.001139935222	&\phantom{a}	&	0.341381192207	&	0.001264145365	\\
0.072	&	-				&	-				&\phantom{a}	&	0.346189767122	&	0.00115209783	&\phantom{a}	&	0.346207857132	&	0.001279652584	\\
0.073	&	-				&	-				&\phantom{a}	&	0.351014554501	&	0.00116395636	&\phantom{a}	&	0.35103482008	&	0.001294906368	\\
0.074	&	-				&	-				&\phantom{a}	&	0.355839431286	&	0.001175460056	&\phantom{a}	&	0.355861902237	&	0.001309957588	\\
0.075	&	-				&	-				&\phantom{a}	&	0.36066454649	&	0.001186609035	&\phantom{a}	&	0.360689252615	&	0.001324755372	\\
0.076	&	-				&	-				&\phantom{a}	&	0.365489840508	&	0.001197352656	&\phantom{a}	&	0.365516811609	&	0.001339299721	\\
0.077	&	-				&	-				&\phantom{a}	&	0.370315164328	&	0.001207538764	&\phantom{a}	&	0.370344519615	&	0.001353590749	\\
0.078	&	-				&	-				&\phantom{a}	&	0.375140637159	&	0.001217218116	&\phantom{a}	&	0.375172406435	&	0.001367526944	\\
0.079	&	-				&	-				&\phantom{a}	&	0.379966259003	&	0.001226238674	&\phantom{a}	&	0.380000561476	&	0.001381159294	\\
0.08	&	-				&	-				&\phantom{a}	&	0.384792000055	&	0.001234499039	&\phantom{a}	&	0.384828835726	&	0.001394436695	\\
0.081	&	-				&	-				&\phantom{a}	&	0.389617711306	&	0.001241948688	&\phantom{a}	&	0.389657229185	&	0.00140725798	\\
0.082	&	-				&	-				&\phantom{a}	&	0.394443571568	&	0.001248486107	&\phantom{a}	&	0.394485831261	&	0.001419521985	\\
0.083	&	-				&	-				&\phantom{a}	&	0.399269461632	&	0.0012539085	&\phantom{a}	&	0.399314552546	&	0.001431279001	\\
0.084	&	-				&	-				&\phantom{a}	&	-				&	-				&\phantom{a}	&	0.404143542051	&	0.001442377339	\\
0.085	&	-				&	-				&\phantom{a}	&	-				&	-				&\phantom{a}	&	0.408972501755	&	0.001452614204	\\
0.086	&	-				&	-				&\phantom{a}	&	-				&	-				&\phantom{a}	&	0.413801699877	&	0.001462040236	\\
0.087	&	-				&	-				&\phantom{a}	&	-				&	-				&\phantom{a}	&	0.418630868196	&	0.001470199204	\\

    \hline
    \hline
  \end{tabular}
\end{table*}

\end{document}